\documentclass[12pt]{article}
\pdfoutput=1

\usepackage{putex}
\usepackage{graphicx}
\usepackage{caption}
\usepackage{amsmath}
\usepackage{array}
\usepackage{subcaption}
\usepackage{epstopdf}
\usepackage{enumerate}
\usepackage{cite}
\usepackage{youngtab}
\usepackage{tensor}
\usepackage{slashed}
\usepackage[aligntableaux=center]{ytableau}
\usepackage[utf8]{inputenc}
\usepackage{rotating}
\usepackage{bigfoot}
\usepackage[
      colorlinks=true,
      linkcolor=blue,
      urlcolor=blue,
      filecolor=black,
      citecolor=red,
      ]{hyperref}

\newcommand{\abs}[1]{\left\lvert #1 \right\rvert}

\newcommand {\be} {\begin {equation}}
\newcommand {\ee} {\end {equation}}

\newcommand {\bes} {\begin {equation*}}
\newcommand {\ees} {\end {equation*}}

\newcommand{\es}[2] {\begin{equation} \label{#1} \begin{split} #2 \end{split} \end{equation}}

\newcommand{\cA}{{\mathcal A}}
\newcommand{\cB}{{\mathcal B}}
\newcommand{\cC}{{\mathcal C}}
\newcommand{\cD}{{\mathcal D}}
\newcommand{\cE}{{\mathcal E}}

\newcommand{\cG}{{\mathcal G}}

\newcommand{\cK}{{\mathcal K}}

\newcommand{\cN}{{\mathcal N}}
\newcommand{\cO}{{\mathcal O}}

\newcommand{\cR}{{\mathcal R}}
\newcommand{\cS}{{\mathcal S}}
\newcommand{\cT}{{\mathcal T}}
\newcommand{\cW}{{\mathcal W}}
\newcommand{\cM}{{\mathcal M}}

\newcommand{\beq}{\begin{equation}}
\newcommand{\eeq}{\end{equation}}

\def\ie{\begin{equation}\begin{aligned}}
\def\fe{\end{aligned}\end{equation}}

\newcommand{\la}{\langle}
\newcommand{\ra}{\rangle}

\newcommand{\m}{\mu}

\newcommand{\A}{{\alpha}}
\newcommand{\B}{{\beta}}
\newcommand{\D}{{\delta}}

\newcommand{\mZ}{{\mathbb Z}}

\newcommand{\mf}{\mathfrak }

\numberwithin{equation}{section}

\def\<{\langle}
\def\>{\rangle}

\begin{document}

\preprint{PUPT-2852}

\institution{PU}{Joseph Henry Laboratories, Princeton University, Princeton, NJ 08544, USA}
\institution{Exile}{Department of Particle Physics and Astrophysics, Weizmann Institute of Science, Rehovot, Israel}

\title{
$\mathcal{N}=4$ Super-Yang-Mills Correlators at Strong Coupling from String Theory and Localization
}

\authors{Damon J.~Binder,\worksat{\PU} Shai M.~Chester,\worksat{\Exile} Silviu S.~Pufu,\worksat{\PU} and Yifan Wang\worksat{\PU}}

\abstract{
 We compute $1/\lambda$ corrections to the four-point functions of half-BPS operators in $SU(N)$ $\cN=4$ super-Yang-Mills theory at large $N$ and large 't Hooft coupling $\lambda=g_\text{YM}^2 N$ using two methods. Firstly, we relate integrals of these correlators to derivatives of the mass deformed $S^4$ free energy, which was computed at leading order in large $N$ and to all orders in $1/\lambda$ using supersymmetric localization. Secondly, we use AdS/CFT to relate these $1/\lambda$ corrections to higher derivative corrections to supergravity for scattering amplitudes of Kaluza-Klein scalars in IIB string theory on $AdS_5\times S^5$, which in the flat space limit are known from worldsheet calculations. These two methods match at the order corresponding to the tree level $R^4$ interaction in string theory, which provides a precise check of AdS/CFT beyond supergravity, and allow us to derive the holographic correlators to tree level $D^4R^4$ order. Combined with constraints from \cite{Alday:2018pdi}, our results can be used to derive CFT data to one-loop $D^4R^4$ order. Finally, we use AdS/CFT to fix these correlators in the limit where $N$ is taken to be large while $g_{\rm YM}$ is kept fixed.  In this limit, we present a conjecture for the small mass limit of the $S^4$ partition function that includes all instanton corrections and is written in terms of the same Eisenstein series that appear in the study of string theory scattering amplitudes.
}
\date{}

\maketitle

\tableofcontents

\pagebreak

\section{Introduction and Summary}
\label{intro}

The duality between the $SU(N)$ ${\cal N} = 4$ super-Yang-Mills (SYM) theory and type IIB string theory on $AdS_5 \times S^5$ \cite{Maldacena:1997re,Witten:1998qj,Gubser:1998bc} (for a review, see \cite{Aharony:1999ti}) states that when both $N$ and the 't Hooft coupling $\lambda = g_\text{YM}^2 N$ are large, the CFT quantities can be computed using classical 10d type IIB supergravity.  Subleading contributions in $1/\lambda$ come from higher derivative corrections to the supergravity action that are due to integrating out massive string modes, while corrections in $1/N$ come from bulk loop diagrams with internal 10d supergravitons.  In this work, we study the $1/\lambda$ corrections to the four-point functions of single trace $1/2$-BPS conformal primary operators in the SYM theory.  Despite the lack of knowledge of the precise 10d bulk interaction vertices that contribute to the four-point functions (see however \cite{deHaro:2002vk,Policastro:2006vt,Paulos:2008tn,Liu:2013dna} for partial results), we will use a combination of recent techniques to determine the first {\em three} terms in the $1/\lambda$ expansion of these correlation functions.  The first term is just the (two-derivative) supergravity result.  The next two terms, suppressed by $1/\lambda^{3/2}$ and $1/\lambda^{5/2}$ respectively, 
 are corrections that correspond to the  effective $R^4$ and $D^4 R^4$ bulk interaction vertices.

In more detail, the standard $\cN = 4$ SYM Lagrangian is written in terms of a vector multiplet consisting of a gauge field $A_\m$, six scalars $\phi^I$ with  $I = 1, \ldots, 6$, and four Weyl fermions $\lambda^A_\A$ with $A=1,2,3,4$, all transforming in the adjoint representation of the $SU(N)$ gauge group. The theory has 
$PSU(2,2|4)$ superconformal symmetry, whose bosonic subgroup includes the conformal group $SO(4,2)$ and $SU(4)_R$ R-symmetry.  Under the latter, the scalars transform as the ${\bf 6}$, while the fermions transform as the ${\bf 4} + \bar {\bf 4}$.  The $1/2$-BPS single-trace conformal primaries we study here are operators of the schematic form $S_p \propto \tr \phi^p$ with $p = 2, 3, \ldots$, which are scalars of scaling dimension $\Delta=p$ and transform in the $SU(4)_R$ representation $[0\, p\, 0]$ (symmetric traceless product of $p$ ${\bf 6}$'s). These BPS operators are holographically dual to scalar Kaluza-Klein modes on $S^5$ \cite{Kim:1985ez}. The lowest one, $S_2\propto\tr \phi^2$, belongs to the same $\cN=4$ superconformal multiplet as the stress-energy tensor.    The connected correlation functions $\langle S_p S_q S_r S_s \rangle$ have been studied extensively in the supergravity limit (i.e.~at leading order in $1/N$ and $1/\lambda$) starting with \cite{Minwalla:1997ka}, with a general formula proposed in \cite{Rastelli:2017udc,Rastelli:2016nze} that, quite surprisingly, was shown to exhibit 10d conformal symmetry \cite{Caron-Huot:2018kta}. (See, for instance, \cite{Eden:2000bk,Arutyunov:2002fh,Arutyunov:2003ae,Berdichevsky:2007xd,Uruchurtu:2008kp,Uruchurtu:2011wh,Arutyunov:2017dti,Arutyunov:2018tvn} for earlier results.)  In this work we will focus on the particular case $\langle S_2 S_2 S_p S_p \rangle$ beyond the supergravity limit.\footnote{Unfortunately, our methods do not currently allow an extension to $\langle S_p S_q S_r S_s \rangle$, with general $p$, $q$, $r$, $s$.  Nevertheless, knowing $\langle S_2 S_2 S_p S_p \rangle$ allows one to then determine  $\langle S_2 S_2 S_2 S_2 \rangle$  to next order in $1/N$ (relative to the one considered here) and up to the same order in $1/\lambda$ \cite{Alday:2018kkw}.}

The main observation that makes our computation possible is that a set of requirements determine the planar $\langle S_2 S_2 S_p S_p \rangle$ correlator order by order in the $1/\lambda$ expansion almost completely,\footnote{For the stress tensor multiplet four-point function in maximally supersymmetric SCFTs, this was first pointed out in 4d in \cite{Alday:2014tsa} and then in 3d and 6d in \cite{Chester:2018aca} and \cite{Chester:2018dga}, respectively, all following the seminal work \cite{Heemskerk:2009pn}. For other work on Mellin space  holographic correlators in various dimensions see \cite{Rastelli:2017ymc,Zhou:2017zaw,Chester:2018lbz }.} up to only a few numerical coefficients at each order. The superconformal Ward identities imply that the correlator is determined by a single function of the conformal cross ratios \cite{Dolan:2001tt}, whose Mellin transform \cite{Penedones:2010ue,Fitzpatrick:2011ia} we denote by $\cM_p(s,t)$. Using crossing symmetry and the analytic structure of tree-level Witten diagrams in Mellin space \cite{Penedones:2010ue,Fitzpatrick:2011ia,Fitzpatrick:2011hu,Fitzpatrick:2011dm}, we can then expand $\cM_p(s,t)$ as a series in $1/\lambda$:
\es{MellinIntro}{
   \cM_p =&c^{-1}\Big( B_1^1\cM_p^{1}+\lambda^{-\frac32}\left[B_4^4 \cM_p^{4}+B_1^4\cM_p^{1} \right] + \lambda^{-2}\left[B_5^5\cM_p^{5}+B_4^5\cM_p^{4}+B_1^5\cM_p^{1} \right]\\
&+\lambda^{-\frac52}\left[B_{6,1}^6 \cM_p^{6,1}+B_{6,2}^6 \cM_p^{6,2}+B_5^6\cM_p^{5}+B_4^6\cM_p^{4}+B_1^6\cM_p^{1} \right]+O(\lambda^{-3})\Big)+O(c^{-2})\,.
}
In this expression, $c = (N^2 - 1)/4$ is the $c$ anomaly coefficient and ${\cal M}_{p}^{n}$ are functions of the Mellin variables $s$ and $t$ (whose expressions will be given in the next section) which for $n>1$ are simply degree $n-4$ polynomials. (We use an extra index $i = 1, 2, \ldots$ and write $\cM_p^{n, i}$ when several such functions are possible.)  The coefficients $B_{n}^m$ are $p$-dependent numbers which we must fix through other methods. Here and in the rest of the paper we find it convenient to expand the $\langle S_2 S_2 S_p S_p \rangle$ correlator in $1/c$ instead of $1/N$ without loss of any information.  The Ward identity from the conservation of the stress tensor fixes the stress tensor OPE coefficients in the $S_2 \times S_2$ and $S_p \times S_p$ OPEs, which fixes the supergravity coefficient $B_1^m$ (in particular $B_1^m = 0$ for $m>1$).  Thus, to determine  $\langle S_2 S_2 S_p S_p\rangle$  to order $1/\lambda^{5/2}$, we need to calculate $B_4^4$, $B^5_{5}$, $B^5_4$, $B_{6,1}^6$, $B_{6,2}^6$, $B_5^6$, and $B_4^6$ as functions of $p$.

The case $p=2$ is simpler, because additional crossing equations not present for $p>2$ fix $B_{6,2}^6 = B_5^6 =B^5_5= 0$, so only four coefficients ($B_4^4$, $B^5_4$, $B_{6,1}^6$, and $B_4^6$) must be determined in this case. One can proceed with two methods. The first method, used first in this context by Goncalves \cite{Goncalves:2014ffa},  involves the relation between CFT correlators and bulk scattering amplitudes  in the flat space limit   \cite{Penedones:2010ue,Fitzpatrick:2011hu}.  Indeed, in this limit, the Mellin amplitude corresponding to  $\langle S_2 S_2 S_2 S_2 \rangle$  is related to the type IIB closed string four-point scattering amplitude of the massless string states (i.e.~of the gravitons and their superpartners from a space-time perspective).  As an expansion in the string coupling $g_s$, the four-point scattering amplitude of these supergravitons takes the form \cite{Polchinski:1998rr}
 \es{GenusZero}{
  \cA = \cA_0 f(s, t) \qquad
   f(s, t) \equiv  -\frac{stu\, \ell_s^6}{64} \frac{\Gamma( - \frac{\ell_s^2 s}{4 }) \Gamma( - \frac{ \ell_s^2 t}{4}) \Gamma( - \frac{\ell_s^2u}{4 }) }{\Gamma( 1+ \frac{\ell_s^2s}{4}) \Gamma( 1+ \frac{\ell_s^2t}{4 }) \Gamma( 1+ \frac{\ell_s^2 u}{4})}  + O(g_s^2)  \,,
 } 
where $\ell_s$ is the string length, the higher order terms in $g_s$  represent contributions from worldsheets of genus one and higher, the quantity $\cA_0$ is the tree-level supergravity scattering amplitude, and $s$ and $t$ are the usual Mandelstam invariants.  We will mostly restrict our attention to the leading term in $g_s$ written in \eqref{GenusZero} which  corresponds to the leading $1/N$ result in the CFT (for fixed $\lambda$ in the {}'t Hooft limit).  Further expanding the function $f(s, t)$ in $\ell_s$, we have
  \es{A10D}{
  f(s, t) =  \left[1 + \ell_{s}^6 f_{R^4}(s, t) + \ell_{s}^{10} f_{D^4 R^4}(s, t)   +
   \ell_s^{12} f_{D^6 R^4}(s, t) + \cdots\right]+O(g_s^2) \,,
 }
with
  \es{fR4}{
  f_{R^4} (s, t) = \frac{ \zeta(3)}{32}stu\,, \quad f_{D^4R^4}(s,t) =\frac{\zeta(5)}{2^{10}}stu(s^2+t^2+u^2)\,, \quad f_{D^6 R^4}(s, t) = \frac{\zeta(3)^2 (stu)^2}{2^{11}} \,,
 }
etc.  As we will explain, the  first quantity in \eqref{fR4} can be used to determine $B_4^4$ \cite{Goncalves:2014ffa}, the second can be used to determine $B_{6,1}^6$, and the absence of an $\ell_s^8$ term in \eqref{A10D} sets $B_5^5=0$.

The second method we use to fix the coefficients in the $1/\lambda$ expansion \eqref{MellinIntro} relies on results from supersymmetric localization. 
For observables that preserve a certain amount of supersymmetry, supersymmetric localization is a powerful method to compute their expectation values by reducing the the path integral to finite-dimensional integrals.   Pestun \cite{Pestun:2007rz} used it to derive a matrix model that calculates the $S^4$ partition function $Z(m,\lambda)$ of a real-mass deformation of the $\cN = 4$ SYM theory known as the $\cN = 2^*$ theory.  Starting from Pestun's result, we follow \cite{Russo:2013kea} (see also \cite{Russo:2013sba,Buchel:2013id,Russo:2013qaa,Russo:2019lgq}) to evaluate the second mass derivative of the $\cN = 2^*$ partition function (evaluated at zero mass) at leading order in $1/N$ and to all orders in $1/\lambda$.\footnote{Quite interestingly, we find that the $1/\lambda$ expansion of the second mass derivative of the $S^4$ planar free energy of the $\cN = 2^*$ theory exhibits some of the zeta functions that one would expect from the string scattering amplitudes.  However, not all of the expected zeta functions appear---for instance the $\zeta(3)^2$ that appears in the string scattering amplitude at order $D^6 R^4$ is absent from the $S^4$ free energy.}${}^{,}$\footnote{At leading order in large $N$ and large $\lambda$, the $\cN = 2^*$ theory on $S^4$ is dual to the supergravity background constructed in \cite{Bobev:2013cja}.} Following the strategy originally outlined in \cite{Binder:2018yvd} for the maximally supersymmetric 3d ABJM \cite{Aharony:2008ug} theory,\footnote{The main difference between 3d and 4d for this strategy is that in 3d the partition function is a function of three masses, while in 4d it's a function of just one mass, but also of $\lambda$ that couples to the marginal operator in the stress tensor multiplet, which does not appear in 3d.  See however \cite{Gorantis:2017vzz} for a proposal for the $S^4$ partition function of the $\cN=1^*$ theory, which is a function of three masses and $\lambda$. } we then show how to relate derivatives of $Z(m,\lambda)$ to integrated four-point functions $\langle S_2S_2S_2S_2\rangle$ and $\langle S_2S_2P\overline P\rangle$, where $P$ is the complex dimension 3 scalar in the stress tensor multiplet and $\langle S_2S_2P\overline P\rangle$ can be related to $\langle S_2S_2S_2S_2\rangle$ using Ward identities that we derive following \cite{Dolan:2001tt}.\footnote{See also \cite{Belitsky:2014zha} where the relation between all stress tensor multiplet four-point functions was described in an abstract language. It would be interesting to derive our explicit formula relating $\langle S_2S_2P\overline P\rangle$ to $\langle S_2S_2S_2S_2\rangle$ from that approach.} Putting these ingredients together allows us to use localization to determine $B_4^4$, $B^5_4$, and a linear combination of $B_{6,1}^6$ and $B_4^6$, namely $7 B_4^6 + 16 B_{6,1}^6$. 

In sum, to derive the order $1/c$ (or $1/N^2$) four-point function  $\langle S_2 S_2 S_2 S_2 \rangle$  at order $1/\lambda^{3/2}$ and $1/\lambda^{2}$,  we have two distinct methods that agree with one another, which is a nontrivial check of AdS/CFT beyond supergravity. In order to pin down the $1/\lambda^{5/2}$ contribution to the four-point function, we need to combine the results from supersymmetric localization and the input from the 10d  scattering amplitude at order  $D^4 R^4$.

The methods of determining the coefficients in \eqref{MellinIntro} can be generalized to $\langle S_2 S_2 S_p S_p \rangle$ for $p>2$.  In this case, the 10d flat space scattering amplitude determines $B_4^4$, $B_5^5$, and $B_{6,i}^6$ as before, and a generalization of Pestun's supersymmetric localization computation developed in \cite{Gerchkovitz:2016gxx} can be used to determine $B_4^4$, $B_4^5$, as well as a linear combination of $B_4^6$, $B_5^6$, and $B_{6,i}^6$.  Without further input, one cannot thus fully fix all the coefficients, although one can put constraints on their $p$-dependence.  Luckily, in \cite{Alday:2018pdi} it was shown that a well-motivated ansatz for the form of the one-loop Mellin amplitudes imposes additional constraints on the quantities $B_4^6$, $B_5^6$, and $B_{6,1}^6$ that can then be combined with the supersymmetric localization and consistency with the string theory flat space scattering amplitude to determine completely the $1/\lambda^{5/2}$ term in $\langle S_2 S_2 S_p S_p \rangle$ for all $p$. These tree level correlators can then be used to derive CFT data to one-loop $D^4R^4$ order \cite{Alday:2018pdi,Aharony:2016dwx}.

While in \eqref{A10D} we expanded the scattering amplitude in $g_s$ first and afterwards in $\ell_s$, one can contemplate a different expansion where we simply expand $f(s, t)$ in $\ell_s$ and at each order keep track of the full $g_s$ dependence.  In such an expansion, $f(s, t)$ takes the form
 \es{fAnotherExp}{
  f(s, t) = 1+ \ell_{s}^6 \tilde f_{R^4}(s, t) + \ell_{s}^8 \tilde f_{\text{1-loop}}(s, t) + \ell_{s}^{10}\tilde  f_{D^4 R^4}(s, t)  +O(\ell_s^{12})\,,
 }
where $\tilde f_{D^{2n} R^4}(s, t)$ are polynomials in $s$ and $t$ as before, but they now depend non-trivially on $g_s$, or more generally on the complexified string coupling $\tau_s$.  The other functions of $s$ and $t$ appearing in \eqref{fAnotherExp}, such as $\tilde f_\text{1-loop}(s, t)$, are non-analytic functions of $s, t$.  The $\tau_s$ dependence of the analytic terms $\tilde f_{R^4}(s, t)$, $\tilde f_{D^4R^4}(s, t)$ and $\tilde f_{D^6R^4}(s, t)$ are captured by certain $SL(2,\mZ)$ invariants that involve non-holomorphic Eisenstein series \cite{Green:1997as,Green:1998by,Green:1999pu,Green:2005ba}, whose expansion at small $g_s$ give a finite number of perturbative contributions and also infinite non-perturbative contributions from D-instantons.  In the $\cN = 4$ SYM theory, the $\ell_s$ expansion can be mapped to a $1/N$ (or $1/c$) expansion at fixed $g_\text{YM}$, which we will refer to as a `very strong coupling' expansion.  As we will discuss in more detail in Section~\ref{finiteg}, while in this limit we do not have good enough control over the supersymmetric localization computation to be able to use it as input, we can nevertheless use the flat space scattering amplitude in order to extract the first couple of terms in the Mellin representation of the $\cN = 4$ SYM correlators.  From them, we can then extract a prediction for the $S^4$ partition function of the $\cN = 2^*$ theory in the very strong coupling limit, which in the field theory would require summing infinitely many instanton contributions \cite{Nekrasov:2002qd,Nekrasov:2003rj,Losev:1997tp,Moore:1997dj}.

The rest of this paper is organized as follows.  In Section~\ref{FOURPOINTMELLIN}, we review properties of the stress tensor multiplet four-point function in the strong coupling limit, and in particular how the Mellin amplitude corresponding to the $\langle S_2 S_2 S_p S_p \rangle$ correlator can be written at leading order in $1/c$ in terms of a few undetermined coefficients at each order in the $1/\lambda$ expansion.  In Section~\ref{LOCALIZATIONCONSTRAINTS} we find the constraints on these coefficients that are imposed from supersymmetric localization.  In Section~\ref{SYMto10}, we combine the supersymmetric localization constraints with the constraints coming from the flat space scattering amplitudes as well as the conjecture of \cite{Alday:2018pdi} to fully determine $\langle S_2 S_2 S_p S_p \rangle$ in the planar limit up to order $1/\lambda^{5/2}$, and to fix CFT data at both tree and 1-loop level. In Section~\ref{finiteg} we discuss the $\langle S_2 S_2 S_p S_p \rangle$ correlator in the very strong coupling limit of the $\cN = 4$ SYM theory. We conclude with a discussion in Section~\ref{disc}.

\section{Four-point functions at large $N$ and strong coupling}
\label{FOURPOINTMELLIN}

Let us begin by discussing the large $N$, strong coupling expansion of the four-point function $\langle S_2 S_2 S_p S_p \rangle$ introduced in the Introduction.  As in the Introduction, we will use the $c$ anomaly coefficient \cite{Anselmi:1997ys}
 \es{cDef}{
  c = \frac{N^2 - 1}{4}
 } 
as our expansion parameter instead of $N$.

 \subsection{Setup}
 \label{setup}

 The operators $S_p$ that we study are single-trace superconformal primary operators of the $\cN = 4$ SYM theory transforming in the $[0\,p\,0]$ irrep of the R-symmetry group $SU(4)_R$.  For fixed $p$, they can be represented as symmetric traceless tensors with $p$ indices $S_{I_1 \ldots I_p}(\vec{x})$, each index ranging from $1$ to $6$.  In terms of the Lagrangian description of the $\cN = 4$ SYM theory these operators are given by\footnote{Here we use $\tr$ to denote the trace in the fundamental representation of $SU(N)$. The $SU(N)$ generators $T^a$ with $a=1,2,\dots,N^2-1$ are normalized such that $\tr (T^a T^b)=\D^{ab}$.}
 \es{SLag}{
  S_{I_1 \ldots I_p}(\vec{x}) = N_p \left[  \tr \left( \phi_{I_1} \cdots \phi_{I_p} \right) - \text{$SO(6)$ traces} \right] \,,
 }
where $\phi_I$, $I = 1, \ldots, 6$, are the adjoint scalar fields of the $\cN = 4$ vector multiplet.  The normalization constant $N_p$ will be given momentarily.  To avoid writing out $SO(6)$ indices, it is convenient to introduce null polarization vectors $Y^I$, with $Y \cdot Y \equiv Y^I Y_I = 0$, and define
 \es{Spol}{
S_p(\vec x,Y)&\equiv S_{I_1\dots I_p}(\vec x)Y^{I_1}\dots Y^{I_p}\,.
}
Because the operators $S_p$ are $1/2$-BPS conformal primaries, their two-point functions are independent of the coupling constant $g_\text{YM}$ \cite{Baggio:2012rr} and can therefore be computed at $g_\text{YM} = 0$ where the $\phi$ propagator is $\langle \phi_I^a (\vec{x}_1) \phi_J^b (\vec{x}_2) \rangle = \frac{\delta^{ab} \delta_{IJ}}{4 \pi^2 \abs{\vec{x}}^2}$, with $a = 1, \cdots, N^2 - 1$ being color indices.  It is convenient to choose the normalization constant $N_p$ in such a way that the two-point function $\langle S_p S_p \rangle$ takes the simple form
 \es{TwoPoint}{
  \langle S_p(\vec{x}_1, Y_1) S_p(\vec{x}_2, Y_2) \rangle 
   = \frac{Y_{12}^p}{\abs{\vec{x}_{12}}^{2p}} \,,  \qquad Y_{12} \equiv Y_1 \cdot Y_2 \,, \qquad
    \vec{x}_{12} \equiv \vec{x}_1 - \vec{x}_2 \,.
 }
At leading order in the $1/c$ expansion, the choice of $N_p$ is given by
 \es{GotNp}{
  N_p = \frac{(2 \pi)^{p} }{\sqrt{p} (4 c)^{p/4}} \,.
 }

Our primary object of study will be the four-point function  $\langle S_2 S_2 S_p S_p\rangle$. Conformal symmetry and $SU(4)_R$ symmetry constrain it to take the form
 \es{22pp}{
 & \langle S_2(\vec x_1,Y_1) S_2(\vec x_2,Y_2) S_p(\vec x_3,Y_3) S_p(\vec x_4,Y_4) \rangle =\\
  &\quad \frac{Y_{34}^{p-2}}{\vec{x}_{12}^4 \vec{x}_{34}^{2p}}
   \biggl[ 
    {\cal S}^1_p(U,V) Y_{12}^2 Y_{34}^2 +  {\cal S}^2_p(U,V) Y_{13}^2 Y_{24}^2 
    +    {\cal S}^3_p(U,V) Y_{14}^2 Y_{23}^2\\
&   \quad \qquad   + {\cal S}^4_p(U,V) Y_{13} Y_{14} Y_{23} Y_{24} 
      + {\cal S}^5_p (U,V) Y_{12} Y_{14} Y_{23} Y_{34} 
       + {\cal S}^6_p (U,V) Y_{12} Y_{13} Y_{24} Y_{34}
    \biggr]\,,
    }
where the $\cS_p^i$ are functions of the conformal cross-ratios 
\es{UVDefD}{
  U \equiv \frac{x_{12}^2 x_{34}^2}{x_{13}^2 x_{24}^2} \,, \qquad 
  V \equiv \frac{x_{14}^2 x_{23}^2}{x_{13}^2 x_{24}^2} \,.
}
The superconformal Ward identities relating the $\cS_p^i$ to one another \cite{Dolan:2001tt}, given in Appendix~\ref{WARDS}, are identical for any $p$.  Quite remarkably, they can be solved in terms of a single unconstrained function $\cT_p$ of $U, V$ by writing $\mathcal{S}^i_p$ as \cite{Dolan:2001tt}:
\es{redSText}{
\mathcal{S}_p^i(U,V) &= \Theta^i(U,V)\cT_p(U,V)+\cS_{p,\text{free}}^i(U,V)\,,\\
 \Theta^i(U,V) &\equiv \begin{pmatrix} V&UV&U&U(U-V-1)&1-U-V&V(V-U-1) \end{pmatrix}\,.
}
Here, $\cS_{p,\text{free}}^i$ is protected and identical to the free SYM correlator. For instance, for $p=2$ we can use Wick contractions with the normalization \eqref{GotNp} to compute\footnote{Note that the $1/c$ term in \eqref{free2} and \eqref{freep} can alternatively be fixed by considering the stress tensor multiplet channel of the superconformal block decomposition, which contributes to the protected correlator $\cS_{i, p, \text{free}}$ and whose coefficient is determined by Ward identities to be $\lambda^2_\text{stress}={p\over 2c}$. } 
\es{free2}{
\cS_{2,\text{free}}^i(U,V)= \begin{pmatrix} 1&U^2&\frac{U^2}{V^2}&\frac1c\frac{U^2}{V}&\frac1c\frac{U}{V}&\frac1cU \end{pmatrix}
} 
and for $p>2$, we have
\es{freep}{
  \cS_{p,\text{free}}^i(U,V)= 
  \begin{pmatrix}
   1&0 &0 & \frac{p(p-1)}{2c}\frac{U^2}{V}&\frac{p}{2c} \frac{U}{V}&\frac{p}{2c} U 
  \end{pmatrix} + O(1/c^2) \,.
}
From now on, we will focus on $\cT_p(U,V)$. Not only does this function determine $\langle S_2S_2S_pS_p\rangle$, but through the superconformal Ward identities it also (along with $c$) uniquely determines all other correlators related to it through supersymmetry. In the Appendix \ref{WARDS} we discuss these superconformal Ward identities in more detail.

\subsection{Strong coupling expansion at large $N$}
\label{tree}

Since $\cN = 4$ SYM has two parameters, $N$ and $g_\text{YM}$, there are in fact several strong coupling limits we could consider at large $N$.  The one we will focus on for most of the paper is the strong coupling 't Hooft limit, in which we first take $N\to\infty$ (or $c \to \infty$) with $\lambda=g_\text{YM}^2N$ fixed, and then take $\lambda\to\infty$.  The double expansion in $c^{-1}$ and $\lambda^{-\frac12}$ is dual to the Type IIB expansion in $g_s^2\ell_s^8$ (counting supergraviton loops) and $\ell_s^2$ (counting higher derivatives), where AdS/CFT relates the string coupling $g_s$ as
\es{gsAds}{
g_s=\frac{g_\text{YM}^2}{4\pi}\,.
}
All tree level terms then have coefficient $c^{-1}$ but different powers of $\lambda$:  for instance, the $R^4$ and $D^4R^4$ contributions are suppressed by $\lambda^{-\frac32}$ and $\lambda^{-\frac52}$, respectively, relative to the supergravity contribution.  A different strong coupling, large $N$ limit, in which we keep $g_\text{YM}^2$ fixed while taking $N \to \infty$ will be discussed in Section~\ref{finiteg}.

In the 't Hooft strong coupling limit, is it simpler to study the $1/c$ contribution to the function $\cT_p(U,V)$ in Mellin space.  The Mellin transform $\cM_p(s,t)$ of $\cT_p(U,V)$ is defined by \cite{Rastelli:2017udc}:
\es{mellinDefD}{
\cT_p(U,V)=&\int_{-i\infty}^{i\infty}\frac{ds\, dt}{(4\pi i)^2} U^{\frac s2}V^{\frac {u-p-2}{2}} \Gamma\left[2-\frac s2\right]\Gamma\left[p-\frac s2\right]\\
&\times\Gamma^{2}\left[\frac{2+p}{2}-\frac t2\right]\Gamma^{2}\left[\frac{2+p}{2}-\frac {u}{2}\right] \cM_p(s,t)\,,
}
where $u = 2p - s - t$, and where the two integration contours in \eqref{mellinDefD} include all poles of the Gamma functions on one side or the other of the contour.  (These poles are associated with double-trace operators \cite{Mack:2009mi}.)   The reason for considering the Mellin space representation is that, in Mellin space, tree-level Witten diagrams are either polynomial in $s, t$ or have a prescribed set of poles and residues that agree with the poles and residues of the conformal blocks corresponding to the single trace operators exchanged in the Witten diagrams.  At large $s, t$, the growth of the tree-level Mellin amplitude is determined by the number of derivatives in the interaction vertices, and a careful analysis shows that $\cM_p^k$ should grow at most as the $(k-4)$th power of $s$ and $t$.  These requirements  should be supplemented by the crossing symmetry relations
 \es{crossM}{
  \cM_p(s,t) = \cM_p(s,u)\,,\qquad \cM_2(s,t) = \cM_2(u,t) \,,
 }
which follow from interchanging the first and second operators, and, for $p = 2$, the first and third.  All these requirements, namely the analytic structure, growth at infinity, and crossing symmetry of $\cM_p(s, t)$ imply that $\cM_p(s, t)$ can be expanded as in Eq.~\eqref{MellinIntro}, which we reproduce here for the reader's convenience:
\es{MellinMainText}{
   \cM_p =&c^{-1}\Big( B_1^1\cM_p^{1}+\lambda^{-\frac32}\left[B_4^4 \cM_p^{4}+B_1^4\cM_p^{1} \right] + \lambda^{-2}\left[B_5^5\cM_p^{5}+B_4^5\cM_p^{4}+B_1^5\cM_p^{1} \right]\\
&+\lambda^{-\frac52}\left[B_{6,1}^6 \cM_p^{6,1}+B_{6,2}^6 \cM_p^{6,2}+B_5^6\cM_p^{5}+B_4^6\cM_p^{4}+B_1^6\cM_p^{1} \right]+  \cdots \Big)+O(c^{-2})\,,
} 
with \cite{Alday:2014tsa}
\begin{equation}\begin{aligned}\label{SintM}
\cM^1_p=&\frac{1}{(s-2)(t-p)(u-p)}\,,\\
\quad\cM^4_p=&1\,,\quad&\cM^5_p=&s\,,\\
\cM^{6,1}_p=&s^2+t^2+u^2\,,\quad &\cM^{6,2}_p=&s^2\,,\\
\end{aligned}\end{equation}
and so on.   Note that while for $k = 1, 4, 5$ there is a unique new Mellin amplitude $\cM_p^k$ of maximal degree,  for $k\geq 6$ there are multiple such amplitudes that are  linearly independent.

Consequently, in position space the function $\cT_p(U,V)$ takes the form:
 \es{SintExp}{
  \cT_p(U,V) &= c^{-1} \biggl[ B_1^1 \cT^1_p
   + \lambda^{-\frac32}\left( B_4^4 \cT^4_p + B_1^4 \cT^1_p  \right) 
    +\lambda^{-2} \left(B_5^5 \cT^5_p + B_4^5 \cT^4_p+ B_1^5 \cT^1_p  \right) \\
    {}&+ \lambda^{-\frac52}\left( B_{6, 1}^6 \cT^{6, 1}_p  + B_{6, 2}^6 \cT^{6, 2}_p + B_5^6 \cT^5_p 
     + B_4^6 \cT^4_p + B_1^6 \cT^1_p \right) + \cdots 
   \biggr] + O(c^{-2}) \,,
 }
with $\cT_p^k$ determined through \eqref{mellinDefD}.  We have
\es{Sint}{
\cT^1_p =& -\frac18U^p\bar D_{p,p+2,2,2}(U,V)\,,\\
\cT^4_p =&\ U^p\bar D_{p+2,p+2,4,4}(U,V)\,,\\
\cT^5_p =&\ 2U^p\left(2\bar D_{p+2,p+2,4,4}(U,V)-\bar D_{p+2,p+2,5,5}(U,V)\right)\,,\\
\cT^{6,1}_p =&\ 2U^p\left(2(1+U+V)\bar D_{p+3,p+3,5,5}-(4+4p-p^2)\bar D_{p+2,p+2,4,4}(U,V)\right)\,,\\
\cT^{6, 2}_p =&\ 4U^p\left(\bar D_{p+2,p+2,6,6}(U,V)-5\bar D_{p+2,p+2,5,5}(U,V)+4\bar D_{p+2,p+2,4,4}(U,V)\right)\,, 
}
where explicit expressions for $\bar D_{r_1,r_2,r_3,r_4}(U,V)$ \cite{Eden:2000bk} can be derived from e.g.~Appendix C in \cite{Binder:2018yvd}.

The leading coefficient $B^n_1(p)$ terms can be determined by demanding that no unphysical twist 2 operators appear in the $s$--channel conformal block decomposition of $\langle S_2S_2 S_pS_p\rangle$ \cite{Dolan:2001tt}, which imposes a relation between $\cS^i_{p,\text{free}}$ and $\cT_p$ in \eqref{redSText}.\footnote{Twist 2 operators correspond to  terms  linear in $U$ in the small $U$ expansion of $\langle S_2S_2 S_pS_p\rangle$. In the large $N$ limit the only twist 2 operators come from the stress tensor multiplet, whose contribution are captured by the superconformal block given in \cite{Dolan:2001tt}. After subtracting this, we find that both $\mathcal{S}^i_{p,\text{free}}$ and $\mathcal{T}_p^1$ still separately include terms linear in $U$, and demanding their cancellation gives \eqref{B1}.} In our normalization we find
\es{B1}{
B_1^1(p)=\frac{4p}{\Gamma(p-1)}\,,\qquad B_1^n(p)=0\quad\text{for}\quad n>1\,.
}
What remains to be determined are the coefficients $B^n_k$ for $k>1$.  We will do so by imposing various constraints, starting in the next section with constraints coming from supersymmetric localization.

\section{Constraints from supersymmetric localization}
\label{LOCALIZATIONCONSTRAINTS}

In the previous section, we used $\cN=4$ superconformal symmetry to fix the form of the four-point correlator $\la S_2S_2S_pS_p\ra$, and we have presented a $1/\lambda$ expansion of this correlator in the strong coupling limit, with undetermined coefficients $B^{n}_{k}(p)$ (see \eqref{SintExp}).  No knowledge of the $\cN = 4$ SYM Lagrangian was needed.  In this section, we connect the discussion in the previous section with exact computations that use the Lagrangian of the 4d $\cN=4$ SYM theory.

On general grounds, in supersymmetric QFTs, the computation of certain observables preserving certain supercharges can be reduced to finite-dimensional matrix integrals using supersymmetric localization. Here we will restrict our attention to the supersymmetric localization setup in \cite{Pestun:2007rz}  (further expanded in \cite{Gerchkovitz:2016gxx}) that involves placing $\cN = 4$ SYM on a round $S^4$ deformed by mass $m$ and chiral couplings $\tau_p$ while preserving the $\cN=2$ subalgebra  $OSp(2|4)$.\footnote{It would be interesting to consider other backgrounds as well.} The four-point function $\la S_2S_2S_pS_p\ra$ with generic $SO(6)_R$ polarizations and the operators inserted at generic positions does not preserve the supercharge used in the localization setup,\footnote{This is except when the $SO(6)_R$ polarizations are aligned so that $\la S_2 S_2 S_p S_p\ra$ becomes extremal. But the extremal correlator in $\cN=4$ SYM is rather trivial as we will see in Section~\ref{loc}.}
but, as we will explain, certain integrated correlators do, and thus can be extracted from localization. In practice, it is convenient to introduce a generating function for these integrated correlators, which is precisely the deformed $S^4$ partition function $Z(m,\tau_p)$  which we will compute using localization. 
Our strategy will be to first identify which integrated correlators can be computed from localization and relate them to the $\la S_2S_2S_pS_p\ra$ correlator at separated points, and then use the localization result to deduce constraints on the constants $B^n_k(p)$ left undetermined from the previous section.

\subsection{Supersymmetric deformations and integrated correlators  }
\label{integrated}

Since the localization computation only uses $\cN = 2$ SUSY, let us start by reviewing the deformations of a general $\cN=4$ SCFT in  flat space that preserves $\cN=2$ Poincar\'e supersymmetry. 
Note that when viewing $\cN = 4$ SYM theory as an $\cN = 2$ SCFT, the $SU(4)_R$ R-symmetry decomposes into $SU(2)_F \times SU(2)_R \times U(1)_R$ where $SU(2)_F$ is the flavor symmetry and $SU(2)_R \times U(1)_R$ is the $\cN=2$ R-symmetry.  Correspondingly, the $\cN=4$ half-BPS operator $S_2$ splits as 
\ie
\begin{array}{c@{\ \to\ }c@{\ \oplus }c@{ \oplus\ }c@{\ \oplus\ }c  }
	\bf 20' & 
	\bf(1,1)_{\pm 2} 
	& \bf(3,3)_0 
	& \bf(1,1)_{0}
	& \bf(2,2)_{\pm 1} \,,
	\\
	S_2
	& \{\cA_2,\bar \cA_2\}
	&  \cB_{(ab)}^{(\A\B)}
	&   \cC
	&  \{\cD_a^\A,\bar \cD_a^\A\}\,,
\end{array}
\label{n4n2dec}
\fe
where $a,b$ are $SU(2)_R$ doublet indices and $\A,\B$ are $SU(2)_F$ doublet indices. Similarly for general half-BPS operators $S_p$, we have
\ie
\begin{array}{c@{\ \to\ }c@{\ \oplus }c   }
	[0\, p\, 0] & 
	\bf(1,1)_{\pm p} 
	&~\dots\,,
	\\
	S_p
	& \{\cA_p,\bar \cA_p\}
	&~ \dots\,.
\end{array}
\fe
The operators $\cA_p,\bar \cA_p$ are $\cN=2$ Coulomb branch (anti)chiral primaries while $\cB$ denotes the Higgs branch chiral primary of $SU(2)_R$ spin $j_R=1$ that generates a current multiplet for the $SU(2)_F$ flavor symmetry. $\cC$ is the bottom component of the $\cN=2$ stress tensor multiplet and $\cD$ are bottom components of the extra $\cN=2$ supercurrent multiplets that must be present due to $\cN = 4$ SUSY\@.

For any $\cN=2$ SCFT, there are two types of deformations preserving $\cN=2$ Poincar\'e supersymmetry \cite{Cordova:2016xhm}. One comes from the mass deformation associated with the current multiplet, in this case for the $SU(2)_F$ flavor symmetry, which takes the form
\ie
m_{\A\B}\int  d^4 x\,    \tilde Q^2 \cB^{(\A\B)}(x) +{\rm c.c.}
\label{Bdef}
\fe
where $m_{\A\B}$ is the triplet of $\mf{su}(2)_F$ masses.  Here and below we use $\tilde Q$ to denote collectively the 4 antichiral supercharges of $\cN=2$ SUSY\@. While $U(1)_R$ and $SU(2)_F$ are broken by this deformation, the $SU(2)_R$ indices are contracted in a way to preserve this R-symmetry subgroup.   Without loss of generality, we can take the mass parameter to lie within the Cartan generated by $(\sigma_2)_{\A\B} = i\diag\{1, 1\}$ and thus $m_{\A\B} = (\sigma_2)_{\A\B} m$.\footnote{The (symmetric) Pauli matrices with lower indices are related to the usual Pauli matrices by $(\sigma_i)_{\A\B} \equiv (\sigma_i)^\gamma{}_\B \epsilon_{\A\gamma}$ with the convention $\epsilon_{12}=1$. }

The other type of deformations correspond to Coulomb branch chiral primaries $\cA_p$   
\ie
\tau_p\int  d^4 x\,   \tilde Q^4 \cA_p (x)+{\rm c.c.}
\label{Adef}
\fe
In particular for $p=2$, this is the familiar exactly marginal deformation of SYM with $\tau_2=\tau\equiv{\theta\over 2\pi}  +  {4\pi i \over g_{\rm YM}^2}$. For general $p$, this deformation again breaks $U(1)_R$ while preserving $SU(2)_R$.  This deformation always preserves $SU(2)_F$.

Any $\cN=2$  QFT with $SU(2)_R$ symmetry can be put on $S^4$ by coupling it to a background $\cN=2$ supergravity multiplet. The background we are interested in breaks the $SU(2)_R$ symmetry to an $U(1)$ subgroup, which together with the $SO(5)$ isometry and 8 supercharges, furnish the full $OSp(2|4)$ symmetry on $S^4$. The deformations \eqref{Bdef} and \eqref{Adef} are also modified by terms that involve the background fields to preserve SUSY on $S^4$, which we will spell out below. The deformed $S^4$ partition function $Z(m,\tau_p)$ of the $\cN = 4$ SYM theory can then be computed using the supersymmetric localization technique. Taking appropriate derivatives with respect to $m$ and $\tau_p$, we obtain the desired integrated correlators.

\subsubsection{$\cN = 2$ flavor current multiplet and real mass deformation on $S^4$}

The $\cN = 2$ conserved current multiplet for an $U(1)$ subgroup of the $SU(2)_F$ symmetry has the following bosonic primary operators:  a conserved current $j_\mu$, a complex scalar $\Sigma$ with scaling dimension $\Delta_\Sigma =3$ and  a triplet of real scalars $\cB_{(ab)}$ of scaling dimension $\Delta_{\cB}=2$.  Any conserved current multiplet $(j_\mu, \Sigma, \cB_{(ab)}, \ldots)$ can be coupled to a background vector multiplet with bosonic components $(A_\mu, \Phi, D_{(ab)}, \ldots)$.  The supersymmetric mass deformation on $S^4$ is obtained by setting $A_\mu = D_{(12)}= 0$ and $\Phi = m=D_{(11)}r=D_{(22)}r$, where $r$ is the radius of $S^4$. We further require   $m$ to be real in order to have a convergent path integral. In the $\cN=1$ notation in which the $\cN = 4$ SYM theory is a theory of a vector multiplet and three adjoint chiral multiplets $(Z_i, \chi_i)$, $i=1, 2, 3$, with $i = 1, 2$ corresponding to the $\cN = 2$ hypermultiplet,\footnote{The $SU(2)_R$ R-symmetry acts on $\bigl( \begin{smallmatrix} ~~Z_1 \\ -\bar Z_2\end{smallmatrix}\bigr)$ and $\bigl( \begin{smallmatrix} Z_2 \\  \bar Z_1\end{smallmatrix}\bigr)$  as doublets.} 
the resulting mass coupling takes the form
\es{MassDeformation}{
	S_m=\int d^4 x\,  \sqrt{g}	 \left( m \left[ \frac{i}{r}  J +  K  \right] +m^2  L \right)\,,
}
where
\es{JDef}{
	J &\equiv \frac 12 \sum_{i=1}^2  \tr \left[ (Z_i)^2 + (\bar Z_i)^2 \right] \,, \qquad
	K \equiv -\frac12  \sum_{i=1}^2 \tr \left(  \chi_i \sigma_2 \chi_i + \tilde \chi_i \sigma_2 \tilde \chi_i \right)\,, \quad
	L\equiv   \sum_{i=1}^2\tr  |Z_i|^2
	\,.
} 
In terms of the $\cN=2$ operators, $J \propto \cB_{(11)}+\cB_{(22)}$ is a real combination of the Higgs branch chiral operators, $K\propto \Sigma+\bar{\Sigma}$ is a real combination of the $\Delta=3$ scalars in the current multiplet,  and $L=\cC+\cK$ is a combination of the  primary operator in the $\cN=2$ stress tensor multiplet \eqref{n4n2dec} and  the Konishi operator  $\cK$,
\ie
\cC\equiv {1\over 3} \tr \left[ |Z_1|^2+|Z_2|^2-2|Z_3|^2  \right]\,, \quad
\cK\equiv  {2\over 3}\sum_{i=1}^3\tr   |Z_i|^2 \,.
\fe
We can further relate the operators $J$, $K$, and $\cC$  to the ${\cal N} = 4$ operators $S_{IJ}$ and $P_{AB}$ via
\es{JToS}{
	J &= N_J \left[ S_{11} + S_{22}  - S_{44} - S_{55} \right] \,, \\
	\cC &= N_\cC \left[ S_{11} + S_{22}  + S_{33}+ S_{44} - 2S_{55}- 2S_{66} \right] \,, \\
	K &= N_K \left[ P_{11} + P_{22} +  \bar P^{11} + \bar P^{22}  \right]  \,,
}  
where the dimension 2 scalar $S_{IJ}$ with $I,J=1,\dots6$ were discussed in the previous section and the dimension 3 complex scalars $P_{AB}$ with $A,B=1,\dots,4$ are discussed in Appendix \ref{22ppApp}. The normalization constants $N_J$ and $N_K$ are independent of the gauge coupling, so they can be computed in the free limit.  Using $\langle Z_i^a(x) \tilde Z_j^b(0) \rangle = \frac{\delta_{ij} \delta^{ab}}{4 \pi^2 \abs{x}^2}$ and $\langle \chi_i^a(x) \tilde \chi_j^b(0) \rangle     = -\frac{ \sigma_\mu x^\mu \delta_{ij} \delta^{ab}}{ 2 \pi^2 \abs{x}^4}$, where $a, b = 1, \ldots, N^2 -1$ for an $SU(N)$ gauge group, we obtain\footnote{ Here $\sigma_\m=(\vec\sigma,-i)$ are 4d chiral gamma matrices where $\vec \sigma$ denotes the usual Pauli matrices.}
\es{GotNJ}{
	N_K^2 = 8 N_J^2 = 36 N_\cC^2 =\frac{N^2-1}{4 \pi^4} \,.
}

One may worry about the appearance of the (bare) Konishi operator in the mass deformation since it a non-BPS operator and hence receives nontrivial renormalizations. Fortunately for the strong coupling limit we are interested in, the Konishi operator is known to develop a large anomalous dimension and effectively decouples from the operator algebra in the ${1\over \lambda}$ expansion \cite{Arutyunov:2000ku,Gromov:2009zb}. Consequently for our purpose, we may effectively set $L=\cC$ in \eqref{MassDeformation}.

\subsubsection{$\cN = 2$ Coulomb branch chiral primaries  on $S^4$}

In general,  Coulomb branch chiral  primary operators $\cA_p$ in $\cN = 2$ SCFTs are complex scalar operators of dimension $\Delta = p$ which are the bottom components of chiral multiplets.  Under the $SU(2)_R \times U(1)_R$ R-symmetry, they transform with charge $p$ under $U(1)_R$ and they are singlets of $SU(2)_R$. The chiral multiplet contains additional scalar operators: a complex scalar $M_p$ of dimension $\Delta=p+2$, and an $SU(2)_R$ triplet of complex scalars $N_p^{(ab)}$ of dimension $\Delta=p+1$. These operators  couple to background chiral multiplets, giving rise to the following supersymmetric deformation on $S^4$ \cite{Gerchkovitz:2016gxx}
\es{ChiralDeformation}{
	S_{\tau_p}=\tau_p \int d^4 x\,  \sqrt{g}	 \left(M_p(x)-i{p-2\over r} (\sigma_2)_{ab} N_p^{(ab)}(x)+{2(p-2)(p-3)\over r^2} \cA_p(x)   \right)\,.
}
As explained in \cite{Gerchkovitz:2016gxx},  the above integral can be reduced to an insertion of the chiral primary operator $\cA_p$ at the North pole of $S^4$,
\ie
S_{\tau_p} =\tau_p \cA_p(N)
\fe
up to $Q$-exact terms.\footnote{A similar phenomenon is encountered in 3d $\cN=4$ superconformal theories where certain integrated correlators on $S^3$ can be reduced, up to $Q$-exact terms, to integrated correlators in the 1d topological theory studied in \cite{Chester:2014mea,Beem:2016cbd,Dedushenko:2016jxl,Dedushenko:2017avn,Dedushenko:2018icp}.
In that case, the 1d theory is defined on a great $S^1$ within $S^3$.
%	 in the case when this 1d theory is defined on a great $S^1$ within $S^3$.
} 
	Similarly, anti-chiral operators $\bar\cA_p$ with dimension $\Delta = p$ and $U(1)_R$ charge $-p$, can be inserted at the South pole of $S^4$ while preserving supersymmetry:
\ie
  S_{\bar\tau_p} =\bar\tau_p \bar \cA_p(S)
\fe

In the $\cN = 4$ SYM theory, one can identify the chiral primaries $\cA_p$ and $\bar \cA_p$ as $\tr Z_3^p$ and $\tr \bar Z_3^p$, respectively, up to normalization.   More abstractly, they correspond to specific components of the $\cN=4$ BPS operators $S_p(Y)$:
\es{ADef}{
	\cA_p \propto S_p(Y_0) \,, \qquad
	\bar \cA_p \propto S_p(\bar Y_0) \,,
}
where $Y_0 = (0, 0, 1, 0, 0, i)$ and $\bar Y_0 = (0, 0, 1, 0, 0, -i)$.

\subsubsection{Integrated mixed correlators on $S^4$}
Using supersymmetric localization, one can compute the $S^4$ partition function $Z(m, \lambda, \tau_p, \bar \tau_p)$ associated to the deformed action
$
S(m,\tau_p)=S_{\cN=4}+S_m+\sum_p( S_{\tau_p}+  S_{\bar\tau_p})
$
as a function of the 't Hooft coupling $\lambda$ as well as the mass parameter $m$ and complex parameters $(\tau_p, \bar \tau_p)$, with the property that derivatives w.r.t.~$m$ correspond to {\em integrated} insertions of $i (J/r + K)$ (as well as $\cC$), and derivatives w.r.t.~$\tau_p$ ($\bar \tau_p$) correspond to insertions of $\cA_p$ ($\bar\cA_p$) at the North (South) poles.  Let us postpone the derivation of $Z(m, \lambda, \tau_p, \bar \tau_p)$ until the next section, and for now assume that we can compute it, or equivalently that we can compute the ratio 
 \es{Ratio}{
  l_p \equiv  \frac{ \partial^2_m \partial_{\tau_p} \partial_{\bar \tau_p} \log Z }{\partial_{\tau_p} \partial_{\bar \tau_p} \log Z}
   = l_p^{(4)} + l_p^{(3)} \,,
 }
with
\es{fDef}{
	l_p^{(4)} = \frac{\int d^4 \vec{x}_1\,  d^4 \vec{x}_2 \, \sqrt{g(\vec{x}_1)} \sqrt{g(\vec{x}_2)} \left\langle   
		\left(i J(\vec{x}_1) + K(\vec{x}_1) \right) \left( i J(\vec{x}_2) + K(\vec{x}_2) \right) \cA_p(N) \bar\cA_p(S) \right\rangle}
	{\left \langle\cA_p(N) \bar\cA_p(S) \right\rangle } 
}
and 
 \es{lp3Def}{
  l_p^{(3)} =  \frac{2\int d^4 \vec{x}  \, \sqrt{g(\vec{x} )}  \la  
	   \cC(\vec{x}) \cA_p(N) \bar\cA_p(S) \ra}
	{\left \langle\cA_p(N) \bar\cA_p(S) \right\rangle } \,.
 }
The ratio \eqref{Ratio} and consequently the expressions \eqref{fDef}--\eqref{lp3Def} are defined so that they are independent of the normalization of $\cA_p$ and $\bar\cA_p$.  (Here and from now on we set the four-sphere radius $r=1$.)

The 3-point function in \eqref{lp3Def} is easy to evaluate because it involves various components of the $\cN = 4$ chiral operators $S_2$ (see \eqref{JToS}), and such 3-point functions are independent of $g_\text{YM}$ \cite{Baggio:2012rr}.  Thus, it can be computed in the free limit using Wick contractions, and we have, at large $c$, 
  \es{ThreePoint}{
  \frac{  \la  
	   \cC(\vec{x}_1) \cA_p(\vec{x}_2) \bar\cA_p(\vec{x}_3) \ra}
	{\left \langle\cA_p(\vec{x}_2) \bar\cA_p(\vec{x}_3) \right\rangle } = -\frac{p}{6 \pi^2} \frac{\vec{x}_{23}^2}{\vec{x}_{12}^2 \vec{x}_{13}^2 }  + O(1/c) \,.
 }
To obtain the same quantity on a four-sphere of unit radius, we write the $S^4$ metric as $ds^2 = \Omega^{-2}(\vec{x}) d\vec{x}^2$ with $\Omega(\vec{x}) = 1 + \frac{\vec{x}^2}{4}$ and make the replacement
\es{S4Dist}{
	\vec{x}_{ij} \to \Omega(\vec{x}_i)^{-1/2} \Omega(\vec{x}_j)^{-1/2}  \vec{x}_{ij}  
}
in all flat space correlators.  Sending $\vec{x}_2 \to 0$ and $\abs{\vec{x}_3} \to \infty$ (corresponding to the North and South poles, respectively), and using $\sqrt{g(\vec{x})} = \Omega(\vec{x})^{-4}$, we find
  \es{lp3Again}{
  l_p^{(3)} = - \frac{p}{3 \pi^2}  \int d^4 \vec{x}\, \frac{1}{\vec{x}^2 \left( 1 + \frac{\vec{x}^2}{4} \right)^2 } + O(1/c)
   = - \frac{4p}{3} +O(1/c) \,.
 }

Using \eqref{JToS}--\eqref{ADef}, the integrand of $l_p^{(4)}$ can be written in terms of the 4-point functions $\langle S_2S_2S_pS_p\rangle$ in \eqref{22pp} as well as $\langle P \bar P S_pS_p\rangle$, which we give in \eqref{22ppR}.  We find
\es{RatioFlat}{
	\frac{ \left\langle   
		\left(i J(\vec{x}_1) + K(\vec{x}_1) \right) \left(i J(\vec{x}_2) + K(\vec{x}_2) \right) \cA_p(\vec{x}_3) \bar\cA_p(\vec{x}_4) \right\rangle}
	{\left \langle\cA_p(\vec{x}_3) \bar\cA_p(\vec{x}_4) \right\rangle }
	= 4  N_J^2 \left[-  \frac{\cS_p^1}{ \vec{x}_{12}^4} + 8  \frac{\cR_p^1 - \cR^2_p + \cR^3_p}{\vec{x}_{12}^6} \right] \,,
} 
where the $\cR^i_p$ are functions of $(U, V)$ defined in \eqref{22ppR}.  Then, taking $\vec{x}_3 = 0$ and $\abs{\vec{x}_4} \to \infty$ and again using $\sqrt{g(\vec{x})} = \Omega(\vec{x})^{-4}$, we can evaluate \eqref{fDef}:
\es{MixedDerp2}{
	l_p^{(4)}
	= 4 N_J^2  \biggl[ - \tilde I_{2} [\cS_p^1]  +8 \tilde I_{3} [\cR^1_p - \cR^2_p + \cR^3_p] \biggr]   \,,
}
where
\es{tildeIDefp}{
	\tilde I_{\Delta} [{\cal G} ] 
	&=   \lim_{\substack{\abs{\vec{x}_3} \to 0 \\ \abs{\vec{x}_4} \to \infty}} \int d^4 \vec{x}_1 \, d^4 \vec{x}_2 \, \frac{ \left( 1 + \frac{x_1^2}{4} \right)^{\Delta - 4}  \left( 1 + \frac{x_2^2}{4} \right)^{\Delta - 4}    }{\vec{x}_{12}^{2 \Delta}  } {\cal G}(U, V) \,.
}
The eight integrals in \eqref{tildeIDefp} can be reduced to three integrals by using the $SO(4)$ symmetry of $S^4$ (fixing the poles).  Indeed, we can use $SO(4)$ to set $\vec{x}_1 = (r_1, 0, 0, 0)$ and $\vec{x}_2 = (r_2 \cos \theta, r_2 \sin \theta, 0, 0)$ and obtain
\es{tildeIDefAgain2}{
	\tilde I_\Delta [{\cal G} ] 
	= \Vol(S^3) \Vol(S^2) \int dr_1\, dr_2\, d\theta \, r_1^3 r_2^3 \sin^2 \theta \frac{ \left( 1 + \frac{r_1^2}{4} \right)^{\Delta - 4}  \left( 1 + \frac{r_2^2}{4} \right)^{\Delta - 4} }{(r_1^2 + r_2^2 - 2 r_1 r_2 \cos \theta)^{\Delta} } {\cal G}\left( \frac{r_1^2 + r_2^2 - 2 r_1 r_2 \cos \theta}{r_1^2}, \frac{r_2^2}{r_1^2}\right)  \,.
}
To simplify this expression further, we can change variables from $(r_1, r_2)$ to $(r, \rho)$, where $r_1 = 2 \rho$ and $r_2 = 2 r \rho$:
\es{tildeIDefAgain3}{
	\tilde I_\Delta [{\cal G} ] 
	=2^{11 - 2 \Delta} \pi^3 \int dr\, d\rho\, d\theta \, r^3 \rho^7 \sin^2 \theta \frac{ \left( 1 + \rho^2 \right)^{\Delta - 4}  \left( 1 + \rho^2 r^2 \right)^{\Delta - 4} }{\rho^{2 \Delta} (1 + r^2 - 2 r \cos \theta)^{\Delta} } {\cal G}\left(1 + r^2 - 2 r \cos \theta , r^2 \right)  \,.
}
Now $\rho$ only appears in the prefactor and the integral over it can be done analytically.  In the cases of interest $\Delta = 2$ and $\Delta = 3$, Eq.~\eqref{tildeIDefAgain3} becomes
\es{tildeIFinal}{
	\tilde I_2 [{\cal G} ] 
	&= 128 \pi^3 \int dr\,  d\theta \, r^3 \sin^2 \theta \, 
	\frac{1 - r^2 + (1 + r^2) \log r}{(r^2 - 1)^3} \frac{  {\cal G}\left(1 + r^2 - 2 r \cos \theta , r^2 \right)  }{ (1 + r^2 - 2 r \cos \theta)^{2} } \,, \\
	\tilde I_3 [{\cal G} ] 
	&= 32 \pi^3 \int dr\,  d\theta \, r^3 \sin^2 \theta \, 
	\frac{\log r}{r^2 - 1} \frac{  {\cal G}\left(1 + r^2 - 2 r \cos \theta , r^2 \right)  }{ (1 + r^2 - 2 r \cos \theta)^{3} }  \,.
}
Note that the functions $\frac{1 - r^2 + (1 + r^2) \log r}{(r^2 - 1)^3}$ and $ \frac{\log r}{r^2 - 1}$ are continuous at $r=1$ despite the fact that the denominator vanishes there.

Thus, the quantity $l_p^{(4)}$ is given by \eqref{MixedDerp2} with the $\tilde I$ given by \eqref{tildeIFinal}.  One can simplify this expression further using the differential relations \eqref{SSPPward} between ${\cal R}^i_p$ and ${\cal S}_p^i$ given by the $\langle P\overline PS_pS_p\rangle$ Ward identity derived in Appendix \ref{WARDS}. After applying integration by parts, we find, at large $c$:
\es{IntByParts}{
	- \tilde I_{2} [\cS^1_p]  +8 \tilde I_{3} [\cR^1_p - \cR^2_p + \cR^3_p] = 16 \pi^4 I[\cT_p] - \frac{2\pi^4}{c} l_p^{(3)}
}
where
\es{tildeIDef}{
	I[{\cal G}] \equiv
	\frac{4}{ \pi}  \int dr\,  d\theta \, r^3 \sin^2 \theta \, 
	\frac{r^2 - 1 - 2 r^2 \log r}{(r^2 - 1)^2} \frac{  {\cG}\left(1 + r^2 - 2 r \cos \theta , r^2 \right)  }{ (1 + r^2 - 2 r \cos \theta)^{2} } \,,
} 
and note that the 3-point function \eqref{lp3Def} appears here due to boundary terms in the integration by parts.

Plugging this into \eqref{MixedDerp2}, we obtain the final result
\es{fFinal}{
 l_p = l_p^{(3)} +l_p^{(4)}  = 8c I[\cT_p] \,.
}
In the next section, we will calculate the quantity $l_p$ from supersymmetric localization, and using \eqref{fFinal} we will then obtain a constraint on the 4-point function $\langle S_2 S_2 S_pS_p \rangle$.

\subsection{Integrated correlators from localization}
\label{loc}

As shown by Pestun \cite{Pestun:2007rz}, the $S^4$ partition function of the ${\cal N} = 2^*$ theory can be computed using supersymmetric localization through a matrix model that takes the form
\es{N2starMatrixModel}{
	Z(m, \lambda)
	= \int d^{N-1} a\, \left( \prod_{i < j} \frac{(a_i  - a_j)^2 H^2(a_i - a_j)}{H(a_i - a_j - m) H(a_i - a_j + m)} \right)
	e^{-\frac{8 \pi^2 N }{\lambda} \sum_i a_i^2} \abs{Z_\text{inst}}^2 \,,
}
where $H(z)$ is a product of two Barnes G-functions, namely $H(z) = G(1+ z) G(1-z)$.  The quantity $\abs{Z_\text{inst}}^2$ represents the contribution to the localized partition function coming from instantons  located at the North and South poles of $S^4$ \cite{Nekrasov:2002qd,Nekrasov:2003rj,Losev:1997tp,Moore:1997dj}.  Because this quantity is non-perturbative and exponentially small when $g_\text{YM} \to 0$ it can be ignored in the 't Hooft limit \cite{Russo:2013kea}.

As already mentioned, \eqref{N2starMatrixModel} was generalized in  \cite{Gerchkovitz:2016gxx} to also include the sources $\tau_p$ and $\bar \tau_p$ for the $\cN = 2$ Coulomb branch operators inserted at the North and South poles of $S^4$.  To achieve this, we consider
\es{MatrixModel}{
	Z(m, \lambda, \tau_p', \bar \tau_p')
	&= \int d^{N-1} a\, \left( \prod_{i < j} \frac{(a_i  - a_j)^2 H^2(a_i - a_j)}{H(a_i - a_j - m) H(a_i - a_j + m)} \right)
	e^{-\frac{8 \pi^2 N }{\lambda} \sum_i a_i^2} \\
	&{}\times e^{i \sum_p \pi^{p/2} (\tau_p' - \bar \tau_p') \sum_i a_i^p}
	\abs{Z_\text{inst}}^2 \,, 
}
where the quantity  $\abs{Z_\text{inst}}^2$ appearing in this formula may be different from the one appearing in \eqref{N2starMatrixModel}, but it can again be ignored  in the 't Hooft limit, and where the parameters $\tau_p'$ and $\bar \tau_p'$ are related to (but not quite the same as) the sources $\tau_p$ and $\bar \tau_p$ for chiral and anti-chiral operators we wanted to introduce.

As explained in \cite{Gerchkovitz:2016gxx}, the difference between $(\tau_p', \bar \tau_p')$ and $(\tau_p, \bar \tau_p)$ is due to operator mixing on $S^4$.  Indeed, if we were to compute the matrix of two-point functions ${\bf A}_{pq}$ of the operators that couple to $\tau_p'$ and $\bar \tau_p'$, 
\es{MatrixTwo}{
	{\bf A}_{pq} \equiv \frac{\partial^2 \log Z}{\partial \tau_p' \partial \bar \tau_q'} \bigg|_{m = \tau_p' = \bar \tau_p' = 0}\,,
}
we would find that this matrix is not diagonal.  The operator basis for which the two-point function matrix is diagonal differs from the naive choice by mixing with operators whose dimensions are strictly lower by multiplets of two.\footnote{The mixing of operators on the supersymmetric round $S^4$ background here is purely due to the metric. Diffeomorphism invariance requires that only the curvature enters in the mixing relations.}  In the 't Hooft limit, we only need to consider mixing between single trace operators, so we should simply diagonalize ${\bf A}$ using the Gram-Schmidt procedure:  for every $n$, we should find an eigenvector $v_n^p$ of the matrix ${\bf A}$ (obeying ${\bf A}_{pq}v_n^p = \lambda_n v_n^q$) normalized such that $v_n^n = 1$ and $v_n^p = 0$ for $p>n$.  Then, the insertion of $\cA_n(N)$ and $\bar \cA_n(S)$ are realized by $\frac{\partial}{\partial \tau_n} = v_n^p \frac{\partial}{\partial \tau_p'}$ and $\frac{\partial}{\partial \bar \tau_n} = \bar v_n^p \frac{\partial}{\partial \bar \tau_p'}$, respectively, up to normalization.  The quantity $l_n$ is defined such that it is independent of the normalizations of $\cA_n(N)$ and $\bar \cA_n(S)$ and is given by
\es{Gotf}{
	l_n = \frac{{\bf B}_{pq} v_n^q v_n^p}{{\bf A}_{pq} v_n^p v_n^q} \,,
}
where we also defined the matrix ${\bf B}$ as
\es{BDef}{
	{\bf B}_{pq} \equiv \frac{\partial^4 \log Z}{ \partial m^2 \partial \tau_p' \partial \bar \tau_q'} \bigg|_{m = \tau_p' = \bar \tau_p' = 0} \,.
}

Let us now use \eqref{MatrixModel} to compute the matrices ${\bf A}$ and ${\bf B}$ and then extract $l_n$.  Because the operator mixing is only between operators whose dimensions differ by an even number and the dimension of $\cA_p$ is $p$, we can focus on even $p$.  (Extending the following analysis to odd $p$ is straightforward, but we won't perform it here.)  Assuming that in the large $N$ limit the eigenvalues become dense, we have
\es{ZApprox}{
	Z(m,\lambda, \tau_p, \bar \tau_p) = \int d^N a \, e^{- N^2 F(a)} \,,
}
where 
\es{FaApprox}{
	F &\approx \frac 12 \int dx \, dy\, \rho(x) \rho(y)  \log \frac{H(x - y + m)H(x - y - m)}{H^2(x- y) \abs{x-y}^2}
	+ \int dx\, \rho(x) V(x)  \,, \\
	V(x) &\equiv\frac{8\pi^2}{\lambda}x^2+\sum_{p=2}^\infty \frac{8\pi^{p/2+1}}{\lambda_p}x^p\,,
}
 we set $Z_\text{inst} = 1$ in \eqref{MatrixModel}, we defined $\lambda_p=\frac{4\pi N}{\Im \tau_p'}$, and we introduced the eigenvalue density $\rho(x)$ normalized such that 
\es{Norm}{
	\int dx\, \rho(x) = 1 \,.
}
When $N$ is large, the integral \eqref{ZApprox} can be evaluated in the saddle point approximation:  $\log Z \approx -N^2 F$, where $F$ is evaluated on the solution to the saddle point equation
\es{saddle}{
	\int dy\, \rho(y)\left(\frac{1}{x-y}-K(x-y)+\frac12K(x-y+m)+\frac12K(x-y-m)\right)=\frac{8\pi^2}{\lambda}x+\sum_{p=2}^\infty\frac{4p\pi^{p/2+1}}{\lambda_p}x^{p-1}\,,
}
where $K(x) \equiv -H'(x) / H(x)$.

To compute the matrix ${\bf A}$ we should first consider $m=0$, in which case we recognize that the partition function is that of a Hermitian matrix model with a polynomial potential $V(x)$.  As is well-studied in random matrix theory (see \cite{DiFrancesco:1993cyw} for a review), the eigenvalue density that solves \eqref{saddle} when $m=0$ is supported on a compact interval $[-b, b]$ and takes the form\footnote{This was first worked out in \cite{Brezin:1977sv} for the ($p=4$) quartic potential. Our case amounts to a simple generalization.}
\es{rhop}{
	\rho_0(x)=\frac{Q(x)}{2\pi}\sqrt{b^2-x^2}\,,
}
where $Q(x)$ is a degree $p-2$ polynomial determined by the requirement that 
\es{support}{
	\frac{V'(x)}{\sqrt{x^2-b^2}}- Q(x)\to \frac{2}{x^2}\,, \qquad \text{as $x \to \infty$} \,.
}
Note that $b$ is also fixed in terms of $\lambda,\lambda_p$ in $V(x)$ by the normalization condition \eqref{Norm}.

The value of $\log Z$ is then approximated by minus Eq.~\eqref{FaApprox}, which after using the saddle point equation \eqref{saddle} can be simplified to
\es{F0Simp}{
	\log Z(0, \lambda, \tau_p', \bar \tau_p')&\approx
	-N^24\pi^2\int dx\, \rho_0(x)\left[\frac{x^2}{\lambda}-\frac{\log x}{4\pi^2}+\sum_p \frac{ \pi^{p/2-1} x^p}{\lambda_p}\right] \,.
}
To compute ${\bf A}_{pq}$ defined in \eqref{MatrixTwo}, we only need to keep the two terms with $\lambda_p$ and $\lambda_q$ in the sum over $p$ in the above expression.  We performed this computation for many pairs of such terms and extracted ${\bf A}_{pq}$.  We found
\es{GotA}{
	{\bf A}_{pq} = \frac{2  \Gamma ( \frac{p+1}{2} ) \Gamma( \frac{q+1}{2} ) }{\pi (p + q) \Gamma(\frac p2 ) \Gamma(\frac q2)} \left( \frac{\lambda}{4 \pi} \right)^{\frac{p+q}{2}}   \,.
}
We can then perform the Gram-Schmidt procedure as described above to find the eigenvectors $v_n^p$ (with even $n$ and even $p$ as above):
\es{vFirstFew}{
	v_2^p &= \begin{pmatrix} 1 & 0  & 0 & 0 & \cdots \end{pmatrix} \,, \\
	v_4^p &= \begin{pmatrix} -\frac{\lambda}{4 \pi}  & 1 & 0 & 0 &  \cdots \end{pmatrix} \,, \\
	v_6^p &= \begin{pmatrix} \frac{9\lambda^2}{256 \pi^2}  & -\frac{3 \lambda}{8 \pi} & 1 & 0 & \cdots \end{pmatrix} \,, \\
}
and so on.  From inspection of these eigenvectors, we find that the general solution for $v_n^p$ is 
\es{vGeneral}{
	v_n^p = (-1)^{n/2} 2^{-n } n  \pi^{ \frac{- 1}{2}} \left( \frac{\lambda}{4 \pi} \right)^{\frac{n-p}{2}} 
	\frac{\Gamma( \frac{1-p}{2} ) \Gamma(\frac{p + n}{2} )  }{ \Gamma(\frac{p+2}{2}) \Gamma( \frac{n - p + 2}{2})} \,,
} 
which implies the following simple result\footnote{This is consistent with fact that the flat space extremal correlators of $\cA_p$ in $\cN=4$ SYM have trivial dependence on the marginal coupling $\tau$ (or $\lambda$) \cite{Lee:1998bxa,DHoker:1998vkc,DHoker:1999jke}.}
	\es{AContracted}{
		v_n^p {\bf A}_{pq} v_n^q = n \left( \frac{\lambda}{16 \pi} \right)^n \,.
	}

We have thus computed the denominator of \eqref{Gotf}.  To compute the numerator, we should first compute the matrix ${\bf B}_{pq}$ defined in \eqref{BDef}.  Because in the expansion of \eqref{FaApprox} at small $m^2$ the leading correction comes at order $m^2$, in order to evaluate this correction we can simply use the $m=0$ eigenvalue density.  Thus,  from \eqref{FaApprox} we find that the second derivative of $\log Z$ with respect to $m^2$ is 
\es{F1}{
	\frac{\partial^2 \log Z}{\partial m^2} \approx 
	N^2\int dx\, dy\, \rho_0(x)\rho_0(y)K'(x-y)\,,
}
where $K(x) = -H'(x) / H(x)$ as defined before.  To evaluate this integral, we make use of the Fourier transform \cite{Russo:2013kea}
\es{Kfour}{
	K'(x)=-\int_0^\infty d\omega \frac{2\omega[\cos(2\omega x)-1]}{\sinh^2\omega}\,,
}
and change variables from $(x, y)$ to $(\xi, \eta)$ given by $\xi =  x/b$ and $\eta =  y/b$.  Then,
\es{F12}{
	\frac{\partial^2 \log Z}{\partial m^2} \approx - \frac{N^2b^4}{2\pi^2}\int_{-1}^1 d\xi d\eta Q(b\xi)Q(b\eta)\sqrt{1-\xi^2}\sqrt{1-\eta^2}\int_0^\infty\frac{\omega\left[\cos[2b\omega(\xi-\eta)]-1\right]}{\sinh^2\omega} \,.
}
Taking into account that $b$ depends on $\tau_p'$ and $\bar \tau_p'$ as given by the normalization condition \eqref{Norm}, we can evaluate \eqref{F12} as a function of $\tau_p'$ and $\bar \tau_p'$ and compute the derivatives required to evaluate ${\bf B}_{pq}$.  The $\xi$ and $\eta$ integrals in \eqref{F12} can be evaluated in terms of Bessel functions.  The computations are very tedious and we were able to perform them only for specific values of $p$.   Quite surprisingly, after using \eqref{vGeneral} and \eqref{AContracted} we find that the quantity $l_n = v_n^p v_n^q {\bf B}_{pq} /(v_n^p v_n^q {\bf A}_{pq}) $ takes the simple form (after renaming $n \to p$)
\es{fnFinal}{
	l_p = 4 p \int_0^\infty d \omega\,  \omega \frac{J_1 ( \frac{\sqrt{\lambda}}{\pi} \omega )^2 - J_p (\frac{\sqrt{\lambda}}{\pi} \omega)^2 }{\sinh^2 \omega} \,.
}
This is our main result that follows from supersymmetric localization.  It can be combined with \eqref{fFinal} to provide a constraint on the four-point function $\langle S_2 S_2 S_p S_p \rangle$.

In the next section, we will be interested in the $1 / \sqrt{\lambda}$ corrections of the correlator $\langle S_2 S_2 S_p S_p \rangle$ at leading order in $1/c$, so let us expand \eqref{fnFinal} in $1/\sqrt{\lambda}$.  Such an expansion can be performed using the Mellin-Barnes representation of the Bessel function (see Appendix~\ref{ASYMPTOTIC}), with the final result
	\es{RatioFinal}{
		l_p &= 2 (p-1) 
		- \frac{4 p (p^2 - 1) \zeta(3)}{\lambda^{3/2}}
		- \frac{3 p (p^2 - 1) (3 - 2 p^2) \zeta(5)}{\lambda^{5/2}} \\
		&{}- \frac{15  p (p^2 - 1) (135 - 124 p^2 + 16 p^4) \zeta(7)}{32 \lambda^{7/2}} \\
		&{}-\frac{35 p (p^2 - 1) (1575 - 1654 p^2+ 320 p^4 -16 p^6 ) \zeta(9)}{64 \lambda^{9/2}} 
		+ \cdots \,.
	}
In the next section, we will use this quantity to fix some of the parameters left undetermined in the expressions \eqref{MellinMainText} and \eqref{SintExp} from Section~\ref{setup}.

\section{Relating $\mathcal{N}=4$ SYM and 10d S-matrix}
\label{SYMto10}

Let us now use various constraints to determine the coefficients $B^n_{k}(p)$ in \eqref{MellinIntro}.  The coefficients $B^n_1(p)$ were already determined in \eqref{B1}, so let us focus on the remaining ones, order by order in the $1 / \lambda$  expansion.  The constraints will come from the comparison of the flat space limit of the CFT correlators with flat space string theory scattering amplitudes, from the supersymmetric localization result \eqref{RatioFinal}, and when $p>2$ also from the implications of a conjecture of \cite{Alday:2018pdi} on the form of one-loop Mellin amplitudes. What we will find is that the reduced Mellin amplitude is
 \es{ReducedMellinFinal}{
  \cM_p(s, t) &= \frac{4p}{\Gamma(p-1)} \frac{1}{c} \biggl[ \frac{1}{(s-2)(t-p)(u-p)} 
   + \frac{(p+1)_3}{4} \zeta(3) \frac{1}{\lambda^{3/2}}   \\
   &{}+ \frac{(p+1)_5}{32} \zeta(5) \left[s^2 + t^2 + u^2 + \frac{2p(p-2)}{p + 5} s
    + \left( -2p^2 + \frac{50 + 20 p (p+2)}{(p+4)(p+5)} \right)  \right] \frac{1}{\lambda^{5/2}} \\
    &{}+ \cdots \biggr]  + O(c^{-2}) \,.
 }
In the particular case $p=2$, this expression simplifies to 
 \es{ReducedMellinTwo}{
    \cM_2(s, t) &= \frac{8}{c} \biggl[ \frac{1}{(s-2)(t-2)(u-2)} 
   + \frac{15 \zeta(3)}{\lambda^{3/2}}   
   + \frac{315 \zeta(5)}{4 \lambda^{5/2}}  \left(s^2 + t^2 + u^2 -3  \right)  
    + \cdots \biggr]  + O(c^{-2}) \,.
 }

\subsection{Constraints from supersymmetric localization}
We will begin with the constraints on $\langle S_2 S_2S_pS_p\rangle $ coming from supersymmetric localization results \eqref{fFinal} and \eqref{RatioFinal}.  The integrals \eqref{tildeIDef} of \eqref{Sint} can be computed numerically to high precision for many values of $p$ to yield:
 \es{integrals}{
  I[\cT_p^1]=&\frac{\Gamma(p)}{ 16p}\,,\qquad I[\cT_p^4]=-\frac{\Gamma(p+2)}{2(p)_4} \,, \qquad
  I[\cT_p^5]=-2p\frac{\Gamma(p+2)}{(p)_5}\,, \\
  I[\cT_p^{6,1}]&=(16 + 20 p - 12 p^2 - 5 p^3 - p^4)\frac{\Gamma(p+2)}{(p)_6}\,, \\
  I[\cT_p^{6,2}]&= -8 p (p-1) \frac{\Gamma(p+2)}{(p)_6} \,,
 }
and so on, where $(p)_n\equiv \Gamma(p+n)/\Gamma(p)$ is the Pochhammer symbol.  Then, we can assemble $l_p$ using \eqref{fFinal} as
 \es{fpGen}{
  l_p &= 8 \biggl[ B_1^1(p) \frac{\Gamma(p)}{ 16p} 
   -  B_4^4(p) \frac{\Gamma(p+2)}{2(p)_4} \frac{1}{\lambda^{\frac 32}}
   - \frac{\left[ 4p B_5^5(p)  + (p+4) B_4^5(p) \right] \Gamma(p+2)}{2(p)_5} \frac{1}{\lambda^2} \\
   &{}+ \left[  \frac{2 B_{6, 1}^6(p) (16 + 20 p - 12 p^2 - 5 p^3 - p^4)  -16 p (p-1) B_{6, 2}^6(p)  - B_5^6(p) 4p (p+5)  - B_4^6(p) (p+4)(p+5)}{2(p)_6}
     \right]\\
     &\times \Gamma(p+2)\frac{1}{\lambda^{\frac 52}} + \cdots
   \biggr] \,.
 }
Comparing with \eqref{RatioFinal}, we see that from the leading term we recover the expression for $B_1^1(p)$ given in \eqref{B1}.  From the subleading terms, we read off:
 \es{CoeffFromLoc}{
  R^4&: \qquad  B_4^4(p) = \zeta(3) \frac{(p)_4}{\Gamma(p-1)} \,, \\
  D^2 R^4 &: \qquad 4p B_5^5(p) + (4 + p) B_4^5(p) = 0 \,, \\
  D^4 R^4 &: \qquad 2 B_{6, 1}^6(p) (16 + 20 p - 12 p^2 - 5 p^3 - p^4)  -16 p (p-1) B_{6, 2}^6(p)  \\
   &\qquad\quad{}- B_5^6(p) 4p (p+5)  - B_4^6(p) (p+4)(p+5) = \frac{3(2p^2 - 3) (p)_6 }{4 \Gamma(p-1)} \zeta(5) \,,  
 }
etc.

\subsection{Constraints from the flat space limit}
For a correlator $\langle \cO_1(\vec{x}_1) \cO_2(\vec{x}_2) \cO_3(\vec{x}_3) \cO_4(\vec{x}_4) \rangle$ of scalar operators in $\mathcal{N}=4$ SYM with the corresponding Mellin amplitude $M^{\cO_1 \cO_2 \cO_3 \cO_4}(s, t)$ (see Appendix~\ref{MELLIN}), we can use \cite{Fitzpatrick:2011hu} to deduce the flat space four supergraviton scattering amplitude with momenta restricted to 5 dimensions (up to an overall numerical factor) \cite{Chester:2018aca,Binder:2018yvd}: 
 \es{FlatFromCFT}{
  {\cal A}(\eta_i, s, t) &=  \Gamma\left(\frac12\Delta_\Sigma-2 \right) \left[ \int_{S^5} d^5x \, \sqrt{g} \prod_{i=1}^4 \Psi_{\eta_i}^{\cO_i} (\vec{n})  \right] \\
  &{}\times
  \lim_{L \to \infty} L^6 \int_{\kappa-i\infty}^{\kappa+ i \infty} \frac{d\alpha}{2 \pi i} \, e^\alpha \alpha^{2 - \frac12\Delta_\Sigma} M^{\cO_1 \cO_2 \cO_3 \cO_4}\left(\frac{L^2}{2 \alpha} s,  \frac{L^2}{2 \alpha} t \right)\,,
 }
where $\Delta_\Sigma$ is the sum of the conformal dimensions of the operators, and the limit is taken at fixed $\ell_s$ and $g_s$ after the CFT quantities $g_\text{YM}$ and $N$ are converted into string theory quantities using the 
AdS/CFT relations
 \es{cTAdS}{
 \frac{L^4}{\ell^4_{s}}= \lambda = {g_\text{YM}^2 N}\,, \qquad
  g_s = \frac{g_\text{YM}^2}{4\pi} \,.
 }
In \eqref{FlatFromCFT}, $\eta_i$ are the polarizations of the supergravitons, and the factor in the square brackets is a form factor involving the wavefunctions $\Psi_{\eta_i}^{\cO_i}(\vec{n})$ of the modes dual to the operators ${\cal O}_i$ in the internal unit $S^5$.  The integration contour in \eqref{FlatFromCFT} must have $\kappa>0$.  It can be seen that at order $g_\text{YM}^{2m} /N^n$ in a double expansion as $N \to\infty$ and $g_\text{YM} \to 0$, only terms that at large $s$ and $t$ scale as $s^a t^b$ with $a+b=2n-3$ contribute to \eqref{FlatFromCFT}, and have coefficient multiplied by $g_s^{m+n} \ell_s^{4n}$. For instance, the leading supergravity term is proportional to $\frac{1}{c}=\frac{4}{N^2-1}$, so in this case $m=0$, $n= 2$, and $a+b=1$, which corresponds to a linear scattering amplitude multiplied by $g_s^2 \ell_s^8$, which is the right scaling for the gravitational constant in 10d.

To apply this formula to $\langle S_2 S_2 S_p S_p \rangle$, we need to compute the Mellin transforms $M_p^i(s,t)$ of the functions $\cS_p^i(U,V)$ in \eqref{22pp}. In Appendix~\ref{MELLIN} we define and compute $M_p^i(s,t)$ in terms of $\cM_p(s,t)$; at leading order in $s, t$ (at each order in the $1/c$ and $1/\lambda$ expansion) we find:
\es{largest}{
 M^i_{p}(s,t) \approx \frac 1 {16}\begin{pmatrix}t^2u^2&s^2u^2&s^2t^2&2s^2tu &2st^2u&2stu^2 \end{pmatrix}\cM_p(s,t)\,,
}
where $u\approx -s-t$ in the large $s,t$ limit. It follows that we can read off the function $f(s, t)$ defined in \eqref{A10D} to get
 \es{Gotfst}{
   f(s, t) = \frac{1}{\cN }\lim_{L \to \infty} L^{14} \int_{\kappa-i\infty}^{\kappa+ i \infty} \frac{d\alpha}{2 \pi i} \, e^\alpha \alpha^{-4-p} {\cal M}_p \left( \frac{L^2}{2 \alpha} s, \frac{L^2}{2 \alpha} t \right) \,,
 }
where the normalization\footnote{We have $ \int \frac{d\alpha}{2 \pi i} e^\alpha \alpha^{-1-p} = \frac{1}{\Gamma(p+1)}$.}  
  \es{GotN}{
  \cN =(4\pi)^2B_1^1(p) \frac{32 g_s^2 \ell_s^8 }{stu} \int \frac{d\alpha}{2 \pi i} e^\alpha \alpha^{-1-p}
   = \frac{2048 \pi^2 g_s^2 \ell_s^8 }{stu} \frac{p}{\Gamma(p-1) \Gamma(p+1)}
 }
is such that the leading term in $f(s, t)$ in the small $\ell_s$ expansion is equal to $1$, which we derive by plugging the leading term ${\cal M}_p = \frac{1}{c} B_1^1(p) {\cal M}_p^1 + \ldots$ in \eqref{MellinMainText} into \eqref{Gotfst}.  (We also used the formula for $B_1^1(p)$ given in \eqref{B1}.)

Using the full expansion \eqref{MellinMainText} in \eqref{Gotfst}, we obtain
 \es{fFromBulk}{
  f(s, t) &= \frac{stu}{B_1^1(p)}\biggl[ \frac{B_1^1(p)}{stu} +  \frac{B_4^4(p) }{2^3 (p+1)_3} \ell_s^6
   +  \frac{B_5^5(p) s}{2^4 (p+1)_4} \ell_s^8\\
    {}&+ \frac{B_{6, 1}^6(p) (s^2 + t^2 + u^2) + B_{6, 2}^6(p) s^2 }{2^5 (p+1)_5} \ell_s^{10} + \cdots \biggr]
     + O(g_s^2) \,.
 }
Comparing with the string theory amplitude \eqref{A10D}, we read off
 \es{CoeffFromFlat}{
  R^4&: \qquad B_4^4(p) = \zeta(3) \frac{(p)_4}{\Gamma(p-1)}  \,, \\
  D^2 R^4&: \qquad B_5^5(p) = 0 \,, \\
  D^4 R^4&: \qquad B_{6, 1}^6(p) =  \zeta(5) \frac{(p)_6}{8 \Gamma(p-1)} \,, \qquad B_{6, 2}^6(p) = 0 \,, 
 }
etc.

Note that the constraint on the coefficient $B_4^4(p)$ of the $R^4$ amplitude agrees between \eqref{CoeffFromLoc} and \eqref{CoeffFromFlat}, thus providing a precision test of AdS/CFT beyond the supergravity approximation!  Note also that \eqref{CoeffFromLoc} and \eqref{CoeffFromFlat} are sufficient to determine $B_5^5 = B_4^5 = 0$, thus showing that there is no $1/c \times 1/\lambda^2$ term in the CFT correlator.

Lastly, note that when $p=2$, we have $B_5^6(2) = 0$, and the constraints \eqref{CoeffFromLoc} and \eqref{CoeffFromFlat} are enough to fix
 \es{p2}{
  B_{6, 1}^6(2) = - \frac{B_4^6(2)}{3} = 630 \zeta(5) \,, \qquad B_{6, 2}^6(2) = 0 \,,
 }
so we have determined the  $\langle S_2 S_2 S_2 S_2 \rangle$  correlator completely up to order $1/c \times 1/\lambda^{5/2}$.

\subsection{Locality on $S^5$}
In the flat space limit both the $AdS_5$ and $S^5$ curvatures tend to zero. We therefore expect that locality of the 10d action will constrain the $p$--dependence of the coefficients $B^{n}_k(p)$.  In the effective action of IIB string theory on $AdS_5\times S^5$, the derivatives associated to interaction vertices (such as $R^4,~D^4R^4$, etc.) are distributed between the $AdS_5$ and internal $S^5$ directions. In terms of the KK mode composition, the derivatives in $S^5$ give rise to polynomials in $p$, much like how derivatives in $AdS_5$ give rise to additional powers of $s$ and $t$. From this, we conclude that
\begin{equation}\label{polyB}
B^{n}_k(p) = C^{n}_k(p)\frac{(p)_k}{\Gamma(p-1)}\,,
\end{equation}
where $C^{n}_k(p)$ is a polynomial of degree $2(n-k)$. The ``kinematic factor'' $\frac{(p)_k}{\Gamma(p-1)}$ arises from the $\alpha$ integral in \eqref{Gotfst}, and is not affected by the additional internal space derivatives that give rise to $C^{n}_k(p)$.

Using $\eqref{polyB}$ and \eqref{CoeffFromFlat}, as well as the fact that $B^6_5(p)$ must vanish when $p=2$, we see that we must determine 7 coefficients to fully fix the $\lambda^{-5/2}$ term of the $\langle SSS_pS_p\rangle$:
\begin{equation}B^6_5(p) = (p-2)(a_1 p + a_0)\frac{\zeta(5)(p)_5}{\Gamma(p-1)}, \ \ \ B^6_4(p) = (b_4p^4+b_3p^3+b_2p^2+b_1p+b_0)\frac{\zeta(5)(p)_4}{\Gamma(p-1)}.\end{equation}
Plugging these expressions into the localization result \eqref{CoeffFromLoc}, we can completely determine $B_4^6(p)$ in terms of $B_5^6(p)$:
\begin{equation}\label{bfroma}\begin{aligned}
&b_0 = \frac{25}4\,,\quad b_1 = 8a_0+5\,,\quad b_2 = 8a_1-4a_0-\frac92\\
&b_3 = -4a_1-\frac 54\,,\quad b_4 =-\frac14\,.
\end{aligned}\end{equation}

\subsection{Constraints from \cite{Alday:2018pdi}}
In \cite{Alday:2018pdi}, two distinct arguments were used to constrain, the order $1/c\times 1/\lambda^{5/2}$ Mellin amplitude. By considering the restrictions on the functional form of the one-loop Mellin amplitude, combined with the requirements coming from the flat space limit of both the tree-level and one-loop Mellin amplitudes, Ref.~\cite{Alday:2018pdi} determined $B_5^6(p)$ and $B_4^6(p)$ up to two unknown constants:\footnote{In the notation of \cite{Alday:2018pdi}, our constants $c_{1,2}$ are instead written as $b_{1,2}^{\text{ABP}} = -2c_{1,2}$, where we have added the superscript ABP to distinguish from our $b_i$'s.}
\es{ABP}{
\qquad B_5^6(p)=\frac 14 p(p-2)\frac{\zeta(5)(p)_5}{\Gamma(p-1)}\,,\quad B_4^6(p)&= -\frac 14(p^4+9p^3+c_2p^2-20p+c_1)\frac{\zeta(5)(p)_4}{\Gamma(p-1)}\,.
}
Comparing to \eqref{bfroma}, we can immediately determine that
\begin{equation}\label{finalAnswer}a_1 = -\frac 14\,,\quad a_0 = 1\,,\quad c_1 = -25\,,\quad  \ \ \ c_2 = 10.\end{equation}
We can also confirm that the expressions for $b_1$, $b_3$, and $b_4$ determined in \cite{Alday:2018pdi} are compatible with both localization and their expression for $B_5^6(p)$.

\subsection{Unprotected CFT data up to order $1/\lambda^{5/2}$}

Now that $\langle S_2 S_2S_pS_p\rangle$ has been fixed to order $c^{-1}\lambda^{-\frac52}$, we can use it to extract any CFT data to this order that we like. For instance, we find the tree level anomalous dimension $\gamma_j\big\vert_{c^{-1}}$ of the unique lowest twist even spin $j$ double trace operators $[S\partial_{\mu_1}\dots\partial_{\mu_j}S]$ to be
\es{anom}{
\gamma_j\big\vert_{c^{-1}}=-\frac{24}{(j+1)(j+6)}-\frac{1}{\lambda^{\frac32}}\frac{4320\zeta(3)}{7}\delta_{j,0}-\frac{\zeta(5)}{\lambda^{\frac52}}\left[{30600}\delta_{j,0}+\frac{201600}{11}\delta_{j,2}\right]+O(\lambda^{-3})\,,
}
where first two terms were computed in \cite{Arutyunov:2000ku,DHoker:1999mic} and \cite{Goncalves:2014ffa}, respectively, and contact terms with $n$-derivatives only contribute to operators up to spin $n/2-4$, as explained in \cite{Heemskerk:2009pn}. For higher twists there are many degenerate double trace operators, so one would need to compute all $\langle S_pS_pS_qS_q\rangle$ correlators to determine their anomalous dimensions \cite{Alday:2017xua,Aprile:2017bgs}.

In \cite{Alday:2018pdi}, the double-discontinuity of the 1-loop amplitude was derived to order $\lambda^{-\frac52}$ in terms of the constants $c_1$ and $c_2$, following a similar derivation for the 1-loop SUGRA term in \cite{Alday:2017xua,Aprile:2017bgs,Aharony:2016dwx}. This double discontinuity can be used to derive CFT data for spins greater than 4 and 6 up to order $\lambda^{-\frac52}$ and $\lambda^{-\frac52}$, respectively. After plugging \eqref{finalAnswer} into the formulae from \cite{Alday:2018pdi}, we find that the 1-loop $R^4$ and $D^4R^4$ corrections to $\gamma_j$ are
\es{R41loop}{
\gamma_j\big\vert_{c^{-2}\lambda^{-\frac32}}=&-103680\frac{ (j+2)_4(j^2+7j+16)(j^2+7j+54)}{(j-4)_6(j+6)_6}\zeta(3)\,, \quad\text{for}  \quad j>4\,,\\
   \gamma_j\big\vert_{c^{-2}\lambda^{-\frac52}}=&-(77j^4(j+7)^4+15452 j^3 (j+7)^3 + 1610364 j^2 (j+7)^2 + 48199536 j (j+7)+401725440)\\
   &\times \frac{1036800(j+2)_4}{(j-6)_8(j+6)_8}\frac{\zeta(5)}{8}
   \,,\quad\text{for}\quad j>6\,,
}
where $c_1,c_2$ were required only to fix $\gamma_j\big\vert_{c^{-2}\lambda^{-\frac52}}$. The 1-loop SUGRA anomalous dimension for $j>0$ can be found in \cite{Alday:2017xua,Aprile:2017bgs}.

\section{Very strong coupling expansion and the Eisenstein series}
\label{finiteg}

So far, we have considered the ${\cal N} = 4$ SYM theory in the strong coupling 't Hooft limit, where one first sends $N \to \infty$ while keeping $\lambda = g_\text{YM}^2 N$ fixed, and then takes $\lambda \to \infty$.  Let us now comment on a different strong coupling expansion, which we will refer to as the ``very strong coupling expansion'' whereby we send $N \to \infty$ while keeping $g_\text{YM}$ fixed.  Thus, there is only one expansion parameter in this limit. 

For simplicity, let us focus on the correlator  $\langle S_2 S_2 S_2 S_2 \rangle$ , thus setting $p=2$ and dropping the $p$ dependence from now on.  In the very strong coupling limit, the reduced Mellin amplitude takes the form
\es{SintExpVerStrong}{
  \cT (U,V) &= \tilde B_1^1(\tau) \cT^1(U, V) \frac{1}{c}  + \tilde B_4^4(\tau) \cT^4(U, V) \frac{1}{c^{\frac 74}} 
   + \cT^{\text{1-loop}}(U, V) \frac{1}{c^2} + \\
   &{}+  \bigl( \tilde B_6^6(\tau) \cT^6(U, V) +  \tilde B_4^6(\tau) \cT^4(U, V) \bigr) \frac{1}{c^{\frac 94}} + \cdots \,.
}
where $\tau =\frac{\theta}{2\pi}+\frac{4\pi i}{g_\text{YM}^2}$ is the complexified gauge coupling.  The functions $\cT^i(U, V)$ are the same as in Eq.~\eqref{Sint} (with $p=2$), and as before in Witten diagram language $\cT^1$ corresponds to the tree-level supergravity contribution while $\cT^{4 + n}$ comes from tree-level diagrams with $D^{2n} R^4$ vertices.  The only difference now is that the coefficients of these interaction vertices have a non-trivial dependence on the dilaton-axion $\tau_s = \chi_s + i g_s^{-1}$.  By the same argument that led to \eqref{B1}, we have
\es{B1VeryStrong}{
 \tilde B_1^1 = B_1^1(2)=8\,.
}

It is hard to obtain a constraint from supersymmetric localization in the very strong coupling limit, because at finite $g_\text{YM}$ the instanton corrections to the $S^4$ partition function that we were previously able to ignore now contribute.  
On the other hand, as we will now show, we can use the known Type IIB amplitude to fix $\tilde B^4_4$ in the CFT\@.  For small $\ell_s$ and fixed $g_s$, the function $f(s, t)$ appearing in the type IIB S-matrix (Eq. \eqref{GenusZero}) takes the form
 \es{A10DVSC}{
   f(s, t) =  1 + \ell_{s}^6 \tilde f_{R^4}(s, t) + \ell_{s}^8 \tilde f_{\text{1-loop}}(s, t) + \ell_{s}^{10}\tilde  f_{D^4 R^4}(s, t)  +O(\ell_s^{12}) \,, 
 }
where each coefficient is now a function of the complexified string coupling $\tau_s=\chi_s+i g_s^{-1}$. The terms $ \tilde f_{R^4}$, $\tilde f_{D^4R^4}$, and $ \tilde f_{D^6R^4}$ are protected, and have been computed exactly in terms of $SL(2,\mathbb{Z})$ invariant functions \cite{Green:1997as,Green:1998by,Green:1999pu,Green:2005ba}. The lower two terms can be written in terms of  the non-holomorphic Eisenstein series $\mathcal{E}_r$ as
\es{Xi}{
\tilde f_{R^4}(s,t)&= \frac{stu}{64}g_s^{{3\over 2}} \mathcal{E}_{3/2}(\tau_s,\bar\tau_s) =\frac{stu}{64}\left[ 2\zeta(3) + {2\pi^2\over 3}g_s^2 +O(e^{-1/g_s}) \right]\,,
\\
\tilde f_{ D^4 R^4}(s,t)&=g_s^{{5\over 2}}  \mathcal{E}_{5/2}(\tau_s,\bar\tau_s) \frac{stu}{2^{11}} (s^2+t^2+u^2) = \frac{stu}{2^{10}} \biggl(  {\zeta(5)} + {2\pi^4\over 135} g_s^4 + {\cal O}(e^{-1/g_s}) \biggr) (s^2+t^2+u^2) \,.
}
Here, the $SL(2,\mathbb{Z})$ Eisenstein series is defined as
\es{eisenstein}{
&\mathcal{E}_{r}(\tau_s,\bar\tau_s) =\sum_{(m,n)\neq (0,0)}\frac{g_s^{-r}}{|m+n\tau_s|^{2r}}\\
&=2\zeta(2r)g_s^{-r}+2\sqrt{\pi}g_s^{r-1}\frac{\Gamma(r-1/2)}{\Gamma(r)}\zeta(2r-1)+\frac{2\pi^r}{\Gamma(r)\sqrt{g_s}}\sum_{m,n\neq0}\abs{\frac{m}{n}}^{r-1/2}K_{r-1/2}(2\pi g_s^{-1}|mn|)e^{2\pi imn\chi_s}\,,
}
where the second line gives the Fourier coefficients of the first, the first two terms are the zero modes, and $K$ is a modified Bessel function. Note that $\chi_s$ only shows up in the $O(e^{-1/g_s})$ terms which are associated to D-instanton contributions in IIB string theory.

Using the flat space limit formula \eqref{FlatFromCFT} and the amplitudes \eqref{Xi}, we can now fix the the $R^4$ coefficient $\tilde B_4^4$ in $\mathcal{N}=4$ SYM for finite $g_\text{YM}$:
\es{B4finiteg}{
\tilde B_4^4=60g_s^{3/2}\mathcal{E}_{3/2}(\tau,\bar\tau)\,.
}
We then relate $\tilde B_4^4$ and $\tilde B_1^1=8$ in \eqref{B1VeryStrong} to $Z(m, \lambda)$ using \eqref{Ratio} and \eqref{integrals} with $p=2$ to get\footnote{Note that from \eqref{AContracted}, $\frac{\partial^2 \log Z}{\partial\tau\partial\bar\tau}\Big\vert_{m=0}={\lambda^2\over 128\pi^2}$.}
  \es{R4Final2}{
   \frac{\partial^4 \log Z}{\partial\tau\partial\bar\tau\partial m^2}\Big\vert_{m=0}= \frac{cg_\text{YM}^4}{16\pi^2}-\frac{3{c^{\frac14}}}{128\sqrt{2}\pi^{7/2}}g_\text{YM}^{4}\mathcal{E}_{3/2}(\tau,\bar\tau)+O(c^0)\,.
 }
We can integrate this relation using the fact that the Eisenstein series $\cE_r$ satisfies the differential equation
\ie
4 g_s^{-2}\partial_{\tau_s}\partial_{\bar\tau_s} \cE_r={r(r-1)}\cE_r\,,
\label{eisendeq}
\fe
and we find 
 	 \es{R4Final3}{
 	 	  \frac{\partial^2 \log Z}{\partial m^2}\Big\vert_{m=0}=  \left[ 8c\log g_\text{YM}-\frac{ \sqrt{2}{c^{\frac14}}}{\pi^{3/2}} \mathcal{E}_{3/2}(\tau,\bar\tau) + O(c^0) \right] +\kappa_1(\tau)+\kappa_2(\bar\tau)\,,
 	 } 	
for arbitrary holomorphic and antiholomorphic functions $\kappa_1(\tau)$ and $\kappa_2(\bar\tau)$, which parameterize the expected ambiguity in $\frac{\partial^2 \log Z}{\partial m^2}\Big\vert_{m=0}$.\footnote{These holomorphic and antiholomorphic ambiguities are generalizations of the K\"ahler ambiguities for the $S^4$ partition of generic 4d $\cN=2$ SCFTs with a conformal manifold \cite{Gomis:2014woa,Gomis:2015yaa}, now in the presence of mass deformations. The relevant supersymmetric counterterms take the following form in the $\cN=2$ chiral superspace \cite{deWit:2010za, Butter:2013lta} 
	\ie
	S_{\rm counter}=\int d^4 x\,  d^4 \theta\, \cE \kappa_1(\tau) W^2\,,
	\fe
	and its complex conjugate. Here $\cE$ is the chiral superspace measure and the rest of the integrand must have Weyl weight $2$ and chiral weight $-2$. In writing the above counterterm, we have promoted $\tau$ to a background chiral multiplet with both vanishing Weyl weight and chiral weight whose bottom component is the marginal coupling, and $W$ is the background vector multiplet of Weyl weight $1$ and chiral weight $-1$ whose bottom component is the mass parameter.  See also \cite{Bobev:2016nua} for a discussion of ambiguities in the $S^4$ free energy of $\cN = 1^*$ theory.
	}  Thus, the Eisenstein series in \eqref{R4Final3} provides a nontrivial prediction for the $O(m^2,c^{\frac{1}{4}})$ term in the mass-deformed SYM free energy, which must come from summing up all the instantons in \eqref{MatrixModel}.\footnote{  We emphasize that the mass deformation is crucial for such nontrivial $\tau$ dependence. In the limit of $m=0$, the $S^4$ free energy simply captures the K\"ahler potential on the conformal manifold of the $\cN=4$ SYM which is rather trivial \cite{Gerchkovitz:2014gta}.}

We can also extract the $R^4$ correction to the tree level anomalous dimension of the lowest twist double trace operator, as we did in the $\lambda\to\infty$ limit in \eqref{anom}. From \eqref{B4finiteg}, we get
\es{anomFiniteg}{
\gamma_j=-\frac 1c\frac{24}{(j+1)(j+6)}-\frac{1}{c^{\frac74}}\frac{135\mathcal{E}_{3/2}(\tau_1,\tau_2)}{7\sqrt{2}\pi^{\frac32}}\delta_{j,0}+O(c^{-2})\,.
}
Note that the $R^4$ contribution in \eqref{anomFiniteg} is the same as in \eqref{anom} except with $\zeta(3)$ replaced with an Eisenstein series according to \eqref{Xi}, and $\lambda$ written in terms of $c$. We conjecture that all the large $\lambda\to\infty$ results from protected terms, such as the $D^4R^4$ term in \eqref{anom} and the 1-loop $R^4$ and $D^4R^4$ terms in \eqref{R41loop}, can be promoted to finite $g_\text{YM}$ expansion in $c^{-1}$ by the rules
\es{replace}{
\lambda\to g_\text{YM}^2 \sqrt{4c + 1} \,,\qquad \zeta(3)\to\frac{g_\text{YM}^{3}}{16\pi^{\frac{3}{2}}}\mathcal{E}_{3/2}(\tau,\bar\tau)\,,\qquad  \zeta(5)\to\frac{g_\text{YM}^{5}}{64\pi^{\frac{5}{2}}}\mathcal{E}_{5/2}(\tau,\bar\tau)\,.
}

\section{Discussion}
\label{disc}

In this paper, we have combined various techniques (CFT Ward identities, supersymmetric localization, matching the flat space limit of CFT correlators with string theory scattering amplitudes) to determine the four-point functions of half-BPS operators of the type $\la S_2 S_2 S_p S_p \ra$ in 4d $\cN=4$ SYM in the strong coupling large $N$ limit.  The CFT correlators have a natural expansion in $1/ \lambda$ that maps to higher derivative corrections to type IIB supergravity. These higher derivative couplings are rather complicated and not known in their complete supersymmetric forms, so to determine their contributions to the $\cN = 4$ SYM correlators, we need to resort to other means than a direct computation via Witten diagrams. Fortunately, kinematic constraints from $\cN=4$ superconformal Ward identities and crossing symmetry greatly reduce the number of independent structures that can appear in these correlators, with a priori undetermined coefficients. Our primary tool to constrain these coefficients comes from the localization computation of the SYM on $S^4$, which is an elegant and efficient way to extract certain integrated combinations of the correlators of our interest in the strong coupling limit.

At leading order, we found that the contribution from $R^4$ to $\la S_2 S_2 S_p S_p\ra$ is completely fixed by the localization result. Moreover, this contribution agrees with the IIB string amplitude in the flat space limit and provides a nontrivial check of AdS/CFT to the first nontrivial subleading order in $1/ \lambda$ beyond supergravity.  At the next subleading order of $D^4R^4$, although the current localization results are insufficient to pin down the contribution to the $\la S_2 S_2 S_p S_p\ra$ correlator, from consideration of bulk locality, the IIB string amplitude on flat space, and thanks to the well-motivated ansatz in \cite{Alday:2018pdi}, we have also determined the $\la S_2 S_2 S_p S_p\ra$ correlator  to this order. These correlators could then be used to extract CFT to one-loop $D^4R^4$ order (i.e. $c^{-2}\lambda^{-\frac52}$) following \cite{Alday:2018pdi}.

We also considered a ``very strong coupling'' limit of the correlator $ \la S_2 S_2 S_p S_p\ra$ where we kept $g_{\rm YM}$ finite while taking $N$ to infinity. In this case, instanton contributions to the $S^4$ free energy of $\cN = 4$ SYM theory are no longer suppressed and their contribution is difficult to evaluate. By comparing with the type IIB string amplitude in flat space (known up to genus three), we deduced a proposal for the dependence of the SYM free energy $F(m,\tau)$ on $\tau=\tau_s$ that incorporates non-perturbative contributions. It would be very interesting to directly compute these instanton effects from the large $N$ matrix model. Furthermore, the $\tau_s$ dependence of the proposed SYM free energy take the form of certain non-holomorphic Eisenstein series which are $SL(2,\mZ)$ invariant functions that satisfy Laplace-type differential equations with respect to $\tau_s$ \eqref{eisendeq}.\footnote{
%	The Eisenstein series $\cE_r$ satisfies the differential equation
%\ie
%4 g_s^2\partial_{\tau_s}\partial_{\bar\tau_s} \cE_r={r(r-1)}\cE_r\,.
%\fe
Such differential equations for the higher derivative couplings $R^4$ and $D^4R^4$ are a consequence of IIB supersymmetry \cite{Green:1998by,Wang:2015jna}.}
   It would be interesting to derive these differential constraints directly from CFT considerations. Together with $SL(2,\mZ)$ invariance of the SYM, we would then be able to prove the full $\tau_s$ dependence we conjectured in \eqref{R4Final2} or \eqref{R4Final3}.

The localization computation we pursued in this paper focused on the $S^4$ free energy $F(m,\tau_p)$ of the $\cN=4$ SYM deformed by a real mass $m$ and chiral couplings $\tau_p$, in the strong coupling expansion. We extracted the integrated correlators by taking two derivatives of $m$ and two of $\tau_p$ ($\bar\tau_p$). Another constraint could come from taking four derivatives of $m$, for which we would need to expand $F(m,\tau_p)$ to quartic order in $m$ in the strong coupling expansion, extending the quadratic calculation of \cite{Russo:2013kea}. A further constraint could come from considering the free energy $F(b)$ on the squashed sphere $S^4_b$, which has also been computed from localization in \cite{Hama:2012bg} and is a nontrivial function of $\lambda$. Derivatives of $b$ would give integrated correlators of the stress tensor multiplet, just as derivatives of $m$ and $\tau_p$ did.\footnote{One could also consider taking various mass and gauge coupling derivatives of the $\cN=1^*$ partition function on $S^4$, for which an expression was proposed in \cite{Gorantis:2017vzz}.} By combining all these localization constraints, one could hope to derive the holographic correlator up to $D^6R^4$ order, which is the highest order protected by supersymmetry, by CFT methods alone, which could then be compared with the string theory prediction to check AdS/CFT to unprecedented precision. To go beyond this order one would require CFT data that is not protected by supersymmetry, perhaps from the numerical bootstrap \cite{Beem:2013qxa,Beem:2016wfs}.\footnote{In 3d, first steps of this kind were taken for the maximally supersymmetric ABJM theory, where bootstrap bounds \cite{Chester:2014fya,Chester:2014mea} have been successfully matched to both localization \cite{Agmon:2017xes} and large $N$ Mellin space calculations \cite{Zhou:2017zaw,Chester:2018lbz}.}

\section*{Acknowledgments} 

We thank Ofer Aharony, Gabriel Cuomo, and Igor Klebanov for useful discussions.  DJB, SSP, and YW are supported in part by the Simons Foundation Grant No.~488653. DJB is also supported in part by the General Sir John Monash Foundation.  SMC is supported by the Zuckerman STEM Leadership Fellowship. SSP is also supported in part by the US NSF under Grant No.~1820651, and by an Alfred P.~Sloan Research Fellowship.  YW is also supported in part by the US NSF under Grant No.~PHY-1620059.

\appendix

\section{Stress tensor multiplet four-point functions}
\label{22ppApp}

In Table \ref{stressTable}, we list the scaling dimensions of all the operators in the stress tensor multiplet, as well as their transformations under the $SO(4)$ Euclidean Lorentz symmetry, the $SU(4)_R$ R-symmetry, and the $U(1)_B$ bonus symmetry under which 4-point functions of half-BPS multiplets in $\mathcal{N}=4$ SCFTs are invariant \cite{Intriligator:1999ff,Intriligator:1998ig}. In addition to the superconformal primary $S$ discussed in the main text, we will make use of the dimension $3$ complex scalar $P$ ($\overline P$) in the $[0\,0\,2]$ ($[2\,0\,0]$), the dimension $\frac52$ complex fermion $\chi^{\alpha}$ ($\overline\chi^{\dot\alpha}$) in the $[0\,1\,1]$ ($[1\,1\,0]$), the dimension 3 complex two-form $F^{\alpha\beta}$ ($\overline F^{\dot\alpha\dot\beta}$) in the $[0\,1\,0]$, and the dimension 3 $R$-symmetry current $j^\mu$ in the adjoint $[1\,0\,1]$. 

\begin{table}
\hspace{-.4in}
\begin{tabular}{c||c|c|c|c|c|c|c|c|c}
 & ${S}$ & $\chi\,,\overline\chi$ & ${P\,,\overline P}$ & $j$ & $F\,,\overline F$& $\Psi\,,\overline\Psi$ & $T$ & $\lambda\,,\overline\lambda$ & ${\Phi}\,,\overline\Phi$\\
 \hline 
  $\Delta$   & $2$ & $\frac52$& $3$& $3$ & $3$& $\frac72$& $4$&  $\frac72$ & $4$  \\
 \hline
Spin $[j,j']$ &$[0,0]$ & $[\frac12,0]\,,[0,\frac12]$& $[0,0]$& $[\frac12,\frac12]$ & $[1,0]\,,[0,1]$& $[1,\frac12]\,,[\frac12,1]$ & $[1,1]$ & $[\frac12,0]\,,[0,\frac12]$ & $[0,0]$ \\
 \hline
${SU}(4)_R$  & ${\bf20'}$ & ${\bf20},{\bf\overline{20}}$ & $\overline{\bf {10}}\,,{\bf 10}$ & ${\bf15}$ & ${\bf6}$& ${\bf4},{\bf \bar 4}$& ${\bf1}$& $\bar{\bf4},{\bf4}$ & ${\bf1}$\\
  \hline
  ${U}(1)_B$ & $0$ & $\frac12\,,-\frac12$ & $1\,,-1$ & $0$ & $1\,,-1$& ${\frac12}\,,-\frac12$& ${0}$& ${\frac32},-\frac32$& ${2},-2$\\
\end{tabular}
\caption{Operators in the $\cN = 4$ stress energy tensor multiplet and their scaling dimensions $\Delta$, spins $[j,j']$ of the Euclidean Lorentz group $SO(4)\cong SU(2)\times SU(2)$, irreps of the $R$-symmetry group $SU(4)_R$, and charges of the bonus symmetry group $U(1)_B$.}
\label{stressTable}
\end{table}

We can write $P$ ($\overline P$) as symmetric tensors $P_{AB}(\vec x)$ ($\overline P^{AB}(\vec x)$), where upper (lower) $A,B$  transform in the $\bf4$ ($\bf{\bar4}$) of $SU(4)_R$. We can contract $\bf4$ ($\bf{\bar4}$) indices with polarization spinors $X_A$ ($\overline X^A$) that satisfy the constraint $X_A \overline X^A=0$, so that
\es{Ppol}{
P(\vec x,\overline X)\equiv P_{AB}(\vec x)\overline X^{A} \overline X^{B}\,,\qquad \overline P(\vec x,X)\equiv \overline P^{AB}(\vec x)X_{A} X_{B}\,.
}
In the index free language, we denote contraction via $\delta_{A}{}^{B}$ by $\cdot$\,. In addition to $\delta_{I}{}^{J}$ and $\delta_{A}{}^{B}$, we can construct the invariant tensors $C^I{}^{AB}$ and $\overline C^I{}_{AB}$ \cite{Bobev:2013cja}:
\es{Cmat}{
C_1=&\begin{pmatrix} 0 & \sigma_1 \\ -\sigma_1 & 0 \\\end{pmatrix}\,,\qquad\quad\;\; C_2=\begin{pmatrix} 0 & -\sigma_3 \\ \sigma_3 & 0 \\\end{pmatrix}\,,\qquad\quad C_3=\begin{pmatrix} i\sigma_2 & 0 \\ 0 & i\sigma_2 \\\end{pmatrix}\,,\\ 
C_4=&-i\begin{pmatrix} 0 & i\sigma_2 \\ i\sigma_2 & 0 \\\end{pmatrix}\,,\qquad C_5=-i\begin{pmatrix} 0 & I_2 \\ -I_2 & 0 \\\end{pmatrix}\,,\qquad C_6=-i\begin{pmatrix} -i\sigma_2 & 0 \\ 0 & i\sigma_2 \\\end{pmatrix}\,,\\ 
}
where $\sigma_i$ are the Pauli matrices and $\overline C_{IAB}$ is the complex conjugate of $ C_{I}^{AB}$.

We can use these invariants to define a product $\wedge$ that relates $Y$, $X$, and $\overline X$ as
\es{notation}{
X\wedge X\to Y^I:=&X_A C^I{}^{AB} X_B\,,\qquad \overline X\wedge \overline X\to Y^I:=\overline X^A \overline C^I{}_{AB} \overline X^B\,,\\
 Y\wedge X\to \overline X^A:=&Y_I C^I{}^{AB} X_B\,,\qquad Y\wedge \overline X\to X_A:=Y_I \overline C^I{}_{AB} \overline X^B\,.\\
}

We can now use these $C^I{}^{AB}$ and $\overline C^I{}_{AB}$ to write the fermions $\chi^{\alpha}$ ($\overline\chi^{\dot\alpha}$) as tensors $\chi^{\alpha}{}^I_A(\vec x)$ ($\overline\chi^{\dot\alpha}{}^{IA}(\vec x)$) with the constraints $\chi^{\alpha}{}^I_A C_I^{AB}=0$ ($\overline\chi^{\dot\alpha}{}^{IA} \overline C_{IAB}=0$). We can then contract with polarizations to get
\es{chipol}{
\chi^\alpha(\vec x,\overline X,Y)\equiv\chi^{\alpha}{}_{IA}(\vec x)Y^I\overline X^A\,,\qquad \overline\chi^{\dot\alpha}(\vec x, X,Y)\equiv\overline\chi^{\dot\alpha}{}_I^{A}(\vec x)Y^I X_A\,,\\
}
where the constraints $\chi^{\alpha}{}_{IA} C^{IAB}=\overline\chi^{\dot\alpha}{}_I^{A} \overline C^I_{AB}=0$ become the requirements that $X\wedge Y=\overline X\wedge Y=0$, which can be satisfied by expressing $X=Y\wedge \overline X'$ and $\overline X=Y\wedge \overline X'$ for some auxiliary spinor polarizations $X'$ and $\overline X'$.

Lastly, $j^\mu$, $F^{\alpha\beta}$, and $\overline F^{\dot\alpha\dot\beta}$ can all be written as tensors with $\bf6$ indices as $j^\mu_{[I_1I_2]}$, $F^{\alpha\beta}_I$, and $\overline F^{\dot\alpha\dot\beta}_I$.\footnote{Since $j^\mu$ transforms in the adjoint of $SU(4)_R$, it could also be written using $\bf4$ and $\bf{\bar4}$ indices as $j^\mu{}_A{}^B$, but we will not use this form in this work.} We then contract with polarizations to define
\es{Jpol}{
 j^\mu(\vec x,Y_1,Y_2)\equiv j^\mu_{[I_1I_2]}(\vec x)Y_1^{I_1}Y_2^{I_2}\,,\quad F^{\alpha\beta}(\vec x,Y)\equiv F^{\alpha\beta}_I(\vec x)Y^I\,,\quad \overline F^{\dot\alpha\dot\beta}(\vec x,Y)\equiv\overline F^{\dot\alpha\dot\beta}_I(\vec x)Y^I\,,
}
where $Y_1,Y_2$ must be explicitly antisymmetrized. We normalize our operators by defining their two-point functions, as we did for $S$ in \eqref{TwoPoint}. For the other operators we use we have
\es{TwoPointOther}{
\langle \chi^\alpha(x_1,\overline X_1,Y_1)\overline\chi^{\dot\alpha}(x_2,X_2,Y_2)\rangle=&\frac{Y_{12}(\overline X_1\cdot X_2)i x_{12}^\mu\sigma^{\alpha\dot\alpha}_\mu}{x_{12}^6}\,,\\
\langle j^\mu(\vec x_1,Y_1,Y_{1'})j^\nu(\vec x_2,Y_2,Y_{2'})\rangle =& \frac{4[Y_{12}Y_{1'2'}-Y_{12'}Y_{1'2})]}{x_{12}^6} \left(\delta^{\mu\nu}-2\frac{x_{12}^\mu x_{12}^\nu}{x_{12}^2}\right)\,,\\
\langle F^{\alpha\beta}(x_1,Y_1)\overline F^{\dot\alpha\dot\beta}(x_2,Y_2)\rangle=&\frac{Y_{12}x_{12}^\mu x_{12}^\nu\sigma^{(\alpha\dot\alpha}_\mu\sigma^{\beta)\dot\beta}_\nu}{x_{12}^8}\,,\\
 \langle P(\vec{x}_1, \overline X_1)  \overline P(\vec{x}_2, X_2) \rangle 
   =& \frac{(\overline X_{1}\cdot X_2)^2}{x_{12}^6} \,.
} 

In the main text we make use of four-point functions $\langle S_2 S_2S_pS_p \rangle$ and $\langle P\overline PS_pS_p \rangle$. The former was already discussed in detail in Section \ref{setup}, while the latter can constrained by conformal symmetry and $SU(4)_R$ symmetry to take the form
  \es{22ppR}{
     \langle P(\vec x_1,\overline X_1) \overline P(\vec x_2,X_2) S_p (\vec x_3,Y_3) S_p (\vec x_4,Y_4) \rangle &=  \frac{Y_{34}^{p-2}}{x_{12}^6 x_{34}^{2p}} \biggl[ 
    {\cal R}_{1,p} (U,V)Y_{34}^2 (\overline X_1 \cdot X_2)^2 \\
    +  {\cal R}_{2,p} (U,V) Y_{34}(\overline X_1 \cdot X_2)  \{X_2, \overline X_1, &Y_3, Y_4\} 
     + {\cal R}_{3,p} (U,V) \{X_2, \overline X_1, Y_3, Y_4\}^2
    \biggr] \,,
 }
 where the $SU(4)_R$ invariant $ \{X,\overline X',Y,Y'\}$ is defined in terms of $C^I{}^{AB}$ and $\overline C^I_{AB}$ introduced above as
\es{invariant}{
 \{X,\overline X',Y,Y'\}=&\frac12X_A \overline X^{\prime B} Y_IY'_J(\overline C^I_{BC}C^J{}^{CA}-\overline C^J_{BC}C^I{}^{CA})\,,
}
 and we normalize $P$ in \eqref{TwoPointOther} so that $\cR_{1,p}(U,V)$ approach $1$ as $U \to 0$. To derive the Ward identities in the next Appendix, we also make use of the four-point functions:
\es{SSCC}{
\langle S_2(\vec x_1,Y_1)&S_2(\vec x_2,Y_2)\chi^\alpha(\vec x_3,\overline X_3,Y_3)\overline \chi^{\dot\beta}(\vec x_4,X_4,Y_4)\rangle \\
= &\frac{i x^\mu_{34}\sigma_\mu^{\alpha\dot\beta}}{x_{12}^4x_{34}^6}\biggr[Y_{12}(\overline X_3\cdot X_4)(Y_{12}Y_{34} \cA_{11}+ Y_{13}Y_{24}\cA_{12}+ Y_{14}Y_{23}\cA_{13}) \\
& + \{X_4,\overline X_3,Y_1,Y_2\}(Y_{12}Y_{34} \cA_{14}+ Y_{13}Y_{24}\cA_{15}+ Y_{14}Y_{23}\cA_{16})\biggr] \\
&+\frac{i\sigma_\mu^{\alpha\dot\beta}}{2x_{12}^6x_{34}^6}\left(x_{24}^2x^\mu_{31}-x_{14}^2x_{32}^\mu+x_{23}^2x_{41}^\mu-x_{13}^2x_{42}^\mu-x_{34}^2x_{21}^\mu-x_{12}^2x_{43}^\mu-2\varepsilon^{\mu\nu\rho\sigma}x_{\nu42} x_{\rho13} x_{\sigma12}\right)\\
&\biggr[Y_{12}(\overline X_3\cdot X_4)(Y_{12}Y_{34}\cA_{21} +Y_{13}Y_{24} \cA_{22}+ Y_{14}Y_{23}\cA_{23}) \\
& + \{X_4,\overline X_3,Y_1,Y_2\}(Y_{12}Y_{34}\cA_{24} + Y_{13}Y_{24}\cA_{25}+Y_{14}Y_{23}\cA_{26} )\biggr],
}
and
\es{SSSJ}{
 \langle S_2(\vec x_1,Y_1)&S_2(\vec x_2,Y_2)S_2(\vec x_3,Y_3)j^\m(\vec x_4,Y_4,Y_5)\rangle \\
&= \frac 1 {x_{12}^4x_{34}^4}\left(\frac{x_{24}^\m}{x_{24}^2}-\frac{x_{34}^\m}{x_{34}^2}\right)\biggr[(Y_{14}Y_{25}-Y_{24}Y_{15}) Y_{13}Y_{23} \cW_{11} \\
&+(Y_{14}Y_{25}-Y_{24}Y_{15})  Y_{13}Y_{12} \cW_{12}+(Y_{14}Y_{25}-Y_{24}Y_{15})  Y_{12}Y_{23} \cW_{13}\biggr] \\
&+\frac 1 {x_{12}^4x_{34}^4}\left(\frac{x_{24}^\m}{x_{24}^2}-\frac{x_{14}^\m}{x_{14}^2}\right)\biggr[ (Y_{14}Y_{25}-Y_{24}Y_{15}) Y_{13}Y_{23} \cW_{21} \\
&+ (Y_{14}Y_{25}-Y_{24}Y_{15})  Y_{13}Y_{12} \cW_{22}+(Y_{14}Y_{25}-Y_{24}Y_{15})  Y_{12}Y_{23} \cW_{23}\biggr] \,. \\
}
and 
\es{SPCC}{
\langle S_2(\vec x_1,Y_1)&P(\vec x_2,\overline X_2)\overline\chi^{\dot\alpha}(\vec x_3, X_3,Y_3)\overline \chi^{\dot\beta}(\vec x_4,X_4,Y_4)\rangle \\
= &\frac{x_{14}^2}{x_{12}^6x_{34}^6x_{24}^2}\left((x_{32}^2+x_{42}^2-x_{34}^2)\varepsilon^{\dot\alpha\dot\beta}-4x_{23}^\mu x_{24}^\nu \bar\sigma^{\dot\alpha\dot\beta}_{\mu\nu}\right)\\ &\biggr[Y_{13}Y_{14}(\overline X_2\cdot X_3)(\overline X_2\cdot X_4)\cB_{11} + \{X_3,\overline X_2,Y_1,Y_4\}\{X_4,\overline X_2,Y_1,Y_3\} \cB_{12} \\
&  + Y_{14}(\overline X_2\cdot X_3) \{X_4,\overline X_2,Y_1,Y_3\} \cB_{13}  + Y_{13}(\overline X_2\cdot X_4) \{X_3,\overline X_2,Y_1,Y_4\}  \cB_{14} \biggr] \\
&+\frac{1}{x_{12}^6x_{34}^6}\left((x_{31}^2+x_{41}^2-x_{34}^2)\varepsilon^{\dot\alpha\dot\beta}-4x_{13}^\mu x_{14}^\nu \bar\sigma^{\dot\alpha\dot\beta}_{\mu\nu}\right)\\
&\biggr[Y_{13}Y_{14}(\overline X_2\cdot X_3)(\overline X_2\cdot X_4)\cB_{21} + \{X_3,\overline X_2,Y_1,Y_4\}\{X_4,\overline X_2,Y_1,Y_3\}  \cB_{22}\\
&  + Y_{14}(\overline X_2\cdot X_3) \{X_4,\overline X_2,Y_1,Y_3\}  \cB_{23} +Y_{13}(\overline X_2\cdot X_4) \{X_3,\overline X_2,Y_1,Y_4\}  \cB_{24}\biggr],
}
and
\es{SSPF}{
\langle S_2(\vec x_1,Y_1)&S_2(\vec x_2, Y_2) P(\vec x_3, \overline X_3)\overline F^{\dot\alpha\dot\beta}(\vec x_4,Y_4)\rangle \\
= &\frac{ x_{14}^2x_{23}^\mu x_{43}^\nu+ x_{24}^2x_{31}^\mu x_{41}^\nu+ x_{34}^2x_{12}^\mu x_{42}^\nu}{x_{12}^6x_{34}^8  } \bar\sigma^{\dot\alpha\dot\beta}_{\mu\nu}Y_{12} \{(Y_1\wedge \overline X_3),\overline X_3,Y_2,Y_4\} \cC_{11}\,,
}
where as usual we have fixed their forms using $SU(4)_R$ symmetry and conformal symmetry. The quantity $\bar\sigma_{\mu\nu}^{\dot\alpha\dot\beta}\equiv\frac{1}{2}\epsilon^{\dot\alpha\dot\delta}(\bar\sigma_{\mu\dot\delta\gamma}\sigma_{\nu}^{\gamma\dot\beta}-\bar\sigma_{\nu\dot\delta\gamma}\sigma_{\mu}^{\gamma\dot\beta})$ is anti-self-dual, and we used the conformal structures given in \cite{Cuomo:2017wme}.\footnote{Our expressions differ by some signs and factors of $i$ since we work in Euclidean signature while \cite{Cuomo:2017wme} is in Lorentzian signature.} The expressions \eqref{SSCC}--\eqref{SSSJ} were also used in \cite{Dolan:2001tt}.

\section{Ward identities}
\label{WARDS}

We will now derive the Ward identities that relate $\langle P\overline PS_pS_p\rangle$ to $\langle S_2 S_2S_pS_p\rangle$. These Ward identities are the same as those that relate $\langle P\overline PS_2S_2\rangle$ to $\langle S_2 S_2S_2 S_2\rangle$,\footnote{To show this relation, following \cite{Dolan:2004mu}, it is enough to consider a particular case: Consider an $\mathcal{N}=4$ SCFT with operators $S'_2,P',\dots$ and an $\mathcal{N}=4$ free theory with a free scalar operator $S_1$ in the $[0\,1\,0]$ of $SU(4)_R$. Define $S_2\equiv S'_2$, $P\equiv P'$, and $S_p\equiv S'_2 S_1^{p-2}$, so that $\langle S_2 S_2S_pS_p\rangle=Y_{34}^{p-2}\langle S'_2S'_2S'_2S'_2\rangle$ and $\langle P\overline PS_pS_p\rangle=Y_{34}^{p-2}\langle P'\overline P'S'_2S'_2\rangle$, which proves this relations for this particular theory, and thus in general.} which we can compute using the component field method of \cite{Dolan:2001tt,Binder:2018yvd}. In particular, we will first determine the most general form of these four-point functions that is consistent with conformal symmetry and $R$-symmetry, and then impose invariance under the Poincar\'e SUSY transformations generated by the supercharges $\overline Q^{\dot\alpha}_{ A}$, which automatically implies invariance under the conjugate supercharges  $Q^{\alpha A}$ and the superconformal generators. We define the action of $\overline Q_{\dot\alpha A}$ on the stress tensor multiplet operators using $\bar\delta_{\dot\alpha}(\overline X)$, defined as:
 \es{SUSYvars}{
\bar\delta^{\dot\alpha}(\overline X)S_2(\vec{x},Y) &= \overline\chi^{\dot\alpha}(\vec{x},\overline X\wedge Y,Y) \,, \\
\bar\delta^{\dot\alpha}(\overline X)\chi^\beta(\vec{x},\overline X',Y) &=\frac14\sigma^{\dot\alpha\beta}_{\mu}j^\mu(\vec x,\overline X\wedge \overline X',Y)+2\sigma^{\dot\alpha\beta}_{\mu}\partial^\mu S(\vec x,\overline X\wedge \overline X',Y)\,, \\
\bar\delta^{\dot\alpha}(\overline X)\overline\chi^{\dot\beta}(\vec{x}, X',Y) &=\frac14\epsilon^{\dot\alpha\dot\beta}\overline P(\vec x,X',\overline X\wedge Y)+(\overline X\cdot X')\overline F^{\dot\alpha\dot\beta}(\vec x,Y)\,, \\
\bar\delta^{\dot\alpha}(\overline X)P(\vec{x},\overline X') &=\frac14\sigma^{\dot\alpha\beta}_{\mu}\partial^\mu \chi_\beta(\vec x,\overline X',\overline X\wedge\overline X')\,, \\
&\text{etc.}
}
The action of $ Q^{\alpha A}$ can be found by taking the complex conjugate of \eqref{SUSYvars}, while the supersymmetry variations of the other stress tensor multiplet operators that were omitted in \eqref{SUSYvars} will not be needed in this work. 

We begin by considering the SUSY variation $0=\bar\delta\langle S_2S_2S_2\chi\rangle$, following the original computation in \cite{Dolan:2001tt}. Using the SUSY variations \eqref{SUSYvars}, we can write this equality schematically as
\es{var1}{
0=\bar\delta\langle S_2S_2S_2\chi\rangle=\langle  \overline\chi S_2S_2\chi\rangle+\langle S_2 \overline \chi S_2\chi\rangle+\langle S_2S_2 \overline \chi \chi\rangle+\langle S_2S_2S_2j\rangle+\langle S_2S_2S_2\partial S_2\rangle\,.
}
We then plug in the conformally and $R$-symmetry invariant forms of the four-point functions on the RHS, as given in \eqref{22pp} and the previous Appendix, which yields a large set of differential equations relating these four-point functions. These include relations purely between invariant structures in $\langle S_2S_2S_2S_2\rangle$:
\es{SSSSward}{
\partial_U \cS^4_2(U,V) &= \frac 2 U\cS^4_2(U,V) + \left(\frac 2 U -\partial_U-\partial_V\right)\cS^2_2(U,V)+\left(\frac 2 U + (U-1)\partial_U+V\partial_V\right)\cS^3_2(U,V)\,,\\ 
\partial_V \cS^4_2(U,V) &= -\frac 1 {V}\cS^4_2(U,V) - \frac 1 V\left(2 - U\partial_U  + (1-U) \partial_V\right)\cS^2_2(U,V)-\left(\partial_U+\partial_V\right)\cS^3_2(U,V) \,.
}
Four other relations follow by applying the crossing relations:
\begin{equation}\begin{aligned}
\cS^3_2(U,V) &= \cS^2_2\left(\frac{U}{V},\frac 1 V\right)\,, \qquad &\cS^6_2(U,V) = \cS^4_2\left(\frac{U}{V},\frac1V\right) \,, \\
\cS^2_2(U,V) &= U^2\cS^1_2\left(\frac{1}{U},\frac{V}{U}\right)\,, \qquad &\cS^5_2(U,V) = U^2\cS^4_2\left(\frac{1}{U},\frac{V}{U}\right) \,. \\
\end{aligned}\end{equation}
It is straightforward to check that the Ward identities can in general be solved by \eqref{redSText}, as mentioned in Section~\ref{setup}.

We next extend the computation of \cite{Dolan:2001tt} by computing the SUSY variation $0=\bar\delta\langle S_2S_2P\overline\chi\rangle$, which we write schematically as
\es{var2}{
0=\bar\delta\langle S_2S_2P\overline\chi\rangle=\langle  \overline\chi S_2P\overline\chi\rangle+\langle S_2 \overline \chi P\overline\chi\rangle+\langle S_2S_2 \partial \chi \overline\chi\rangle+\langle S_2S_2P\overline P\rangle+\langle S_2S_2P\overline F\rangle\,.
}
We then plug in the conformally and $R$-symmetry invariant forms of the four-point functions on the RHS, which again yields a large set of differential equations relating these four-point functions. Since $\langle SS\chi\overline \chi\rangle$ was already related by a first order differential equation to $\langle SSSS\rangle$ from \eqref{var1}, the relation between $\langle S_2S_2 \partial \chi \overline\chi\rangle$ and $\langle S_2S_2P\overline P\rangle$ gives a second order differential equation relation between $\langle S_2S_2S_2 S_2\rangle$ and $\langle P\overline P S_2S_2\rangle$, and thus also between $\langle S_2S_2S_p S_p\rangle$ and $\langle P\overline PS_pS_p\rangle$:
\es{SSPPward}{
\cR_{1,p}(U&,V)=\frac18\left[  2U(U-V-3)\partial_U \cS_{1,p}(U,V) +U V (2 - U + 2 V)\partial_V^2 \cS_{1,p}(U,V)\right.\\
&+U^2(U - 2 -2 V)\partial^2_U \cS_{1,p} (U,V) - (4 V^2-4 + U[1 + U - 5 V] ) \partial_V \cS_{1,p}(U,V)\\
&\left. -U(U - 2 -2 V)(U+V-1)\partial_V\partial_U \cS_{1,p}(U,V)+8\cS_{1,p}(U,V)\right]\,,\\
\cR_{2,p}(U&,V)=\frac14\left[ (4V+2UV-2-2V^2 ) \partial_V \cS_{1,p}(U,V)+U V (V-1)\partial_V^2 \cS_{1,p}(U,V)\right.\\
&+U(1+U-V)\partial_U \cS_{1,p}(U,V)+U(V-1)(U+V-1)\partial_V\partial_U \cS_{1,p}(U,V)\\
&\left.+U^2(V-1)\partial^2_U \cS_{1,p}(U,V)\right]\,,\\
\cR_{3,p}(U&,V)=\frac18\left[ U(1+U-V ) \partial_V \cS_{1,p}(U,V)+U^2 V \partial_V^2 \cS_{1,p}\right.\\
&\left.+U^2(U +V-1)\partial_V\partial_U \cS_{1,p}(U,V)+U^3\partial^2_U \cS_{1,p}(U,V)\right]\,.\\
}

\section{Mellin amplitudes}
\label{MELLIN}
In this Appendix we shall define the Mellin transforms $M^i_p(s,t)$ of $\cS^i_p(U,V)$, and then describe how they are related to $\cM_p(U,V).$ To compute the Mellin transform, we first compute the connected correlator:
\begin{equation}\cS^i_{p,\text{conn}}(U,V) \equiv\cS^i_{p} (U,V)-\cS^i_{p,\text{disc}}(U,V)\,\end{equation}
where the disconnected part is given by 
\es{GDisc}{
\cS^i_{2,\text{disc}}(U,V)=& \begin{pmatrix} 1&U^2&\frac{U^2}{V^2} &0&0&0\end{pmatrix}\,,\\
\cS^i_{p,\text{disc}}(U,V)=& \begin{pmatrix} 1&0&0 &0&0&0\end{pmatrix}\quad\text{for}\quad p>2\,.\\
} 
We then define $M^i_p(s,t)$ by the inverse Mellin transform
\es{mellinDef}{
\cS^i_{p,\text{conn}}(U,V)=&\int_{-i\infty}^{i\infty}\frac{ds\, dt}{(4\pi i)^2} U^{\frac s2}V^{\frac {u-p-2}{2}} \Gamma\left[2-\frac s2\right]\Gamma\left[p-\frac s2\right]\\
&\times\Gamma^{2}\left[\frac{2+p}{2}-\frac t2\right]\Gamma^{2}\left[\frac{2+p}{2}-\frac u2\right] M^i_p(s,t)\,,
}
where $u=2p+4-s-t$. Note that although the form of the inverse Mellin transform in \eqref{mellinDef} is identical to that defining $\cM_p(s,t)$ in \eqref{mellinDefD}, the variable $u$ is defined differently. 

Recall from Section~\ref{setup} that for $\langle S_2S_2S_pS_p\rangle$, the Ward identities imply (see Eq.~\eqref{redSText})
\es{redSText2}{
\mathcal{S}_p^i(U,V) &= \Theta^i(U,V)\cT_p(U,V)+\cS_{p,\text{free}}^i(U,V)\,,\\
 \Theta^i(U,V) &\equiv \begin{pmatrix} V&UV&U&U(U-V-1)&1-U-V&V(V-U-1) \end{pmatrix}\,.
}
To compute $M_p^i(s,t)$ we take the Mellin transforms of both sides of this equation. Upon doing so we find that\footnote{Note that $\cS_{i,p,\text{free}}$ vanishes under this Mellin transform, but can be recovered from the inverse Mellin transform of $M_{p,i}(s,t)$ in \eqref{fulltoRed} by carefully regularizing the Mellin amplitude \cite{Rastelli:2017udc}.}
\es{fulltoRed}{
M_{p}^i(s,t)=\widehat\Theta_i(U,V)\circ\mathcal{M}_p\,,
}
where $\widehat\Theta_i(U,V)$ is defined as $\Theta_i(U,V)$ with $U^mV^n$ replaced by a difference operator
\es{thetaMellin}{
\widehat{U^mV^n}\circ{ \cM_p }(s,t)={ \cM_p }(s-2m,t-2n)\left(\frac{4-s}{2}\right)_m^2\left(\frac{4-t}{2}\right)_{2-m-n}^2\left(\frac{4-u}{2}\right)_{n}^2\,,
}
with $u = 2p+4-s-t$ as in \eqref{mellinDef}. In the large $s, t$ limit this simplifies to:
\es{thetaMellinLarge}{
\widehat{U^mV^n}\circ{ \cM_p }(s,t) \underset{s,t\rightarrow\infty}\longrightarrow \frac 1 {16} s^{2m}t^{4-2m-2n}u^{2n}{\cM_p }(s,t)\,,
}
so that we get \eqref{largest} at large $s$ and $t$.

\section{Asymptotic Expansion using Mellin-Barnes Representations}
\label{ASYMPTOTIC}
In this appendix we will explain how to asymptotically expand the integrals such as \eqref{fnFinal}:
\begin{equation}
l_p = 4p\int_0^\infty d\omega\ \omega \frac{J_1\left(\frac{\sqrt\lambda}\pi \omega\right)^2 - J_p\left(\frac{\sqrt\lambda}\pi \omega\right)^2}{\sinh^2(\omega)}
\end{equation}
using a Mellin-Barnes representation of the Bessel function.\footnote{We are gratefully for the \verb|MathOverflow| user \verb|Paul Enta| for bringing this method to our attention in \verb|https://mathoverflow.net/questions/315264/asymptotic-expansion-of-bessel-function-integral|.} Rather than consider $l_p$ itself, for simplicity we will asymptotically expand a streamlined form of the integral
\begin{equation}\label{easybint}
 I_p(x) = \int_0^\infty  d\omega\  \frac{\omega J_p(x \omega)^2}{\sinh^2\omega} \,.
\end{equation}
The first step is to use the Mellin-Barnes representation of products of Bessel functions (see page 436 of \cite{watson1995treatise}):
\begin{equation}\label{mbbessel}
J_\mu(x)J_\nu(x) = \frac 1 {2\pi i}\int_{c-\infty i}^{c+\infty i}\frac{\Gamma(-s)\Gamma(2s+\mu+\nu+1)\left(\frac12 x\right)^{\mu+\nu+2s}}{\Gamma(s+\mu+1)\Gamma(s+\nu+1)\Gamma(s+\mu+\nu+1)} \,,
\end{equation}
which holds for $x>0$, and where $-p+\frac 12<\text{Re}(c)<0$. Substituting \eqref{mbbessel} into \eqref{easybint}, we find that
\begin{equation}
I_p(x) = \int_0^\infty d\omega \int_{-i\infty}^\infty ds\ \frac{\Gamma(-s)\Gamma(2s+2p+1)x^{2p+2s}}{2^{2p+2s}\Gamma(s+p+1)^2\Gamma(s+2p+1)}\frac{\omega^{2p+2s+1}}{\sinh^2\omega}\,.
\end{equation}
We can then perform the integral over $\omega$ explicitly
\begin{equation}
\int_0^\infty d\omega\ \frac{\omega^{2p+2s+1}}{\sinh^2\omega} = \frac{1}{2^{2p+2s}}\Gamma(2p+2s+2)\zeta(2p+2s+1) \,,
\end{equation}
valid for $\text{Re}(s)>-p$. Thus by taking $-p<\text{Re}(c)<0$, we have
 \es{IpGen}{
I_p(x) = \frac 1{2\pi i}\int_{c-i\infty}^{c+i\infty} ds\ \frac{\Gamma(-s)\Gamma(2s+2p+1)\Gamma(2p+2s+2)\zeta(2p+2s+1)}{\Gamma(s+p+1)^2\Gamma(s+2p+1)}\left(\frac{x}{4}\right)^{2p+2s}\,.
 }
This integral can be evaluated asymptotically by closing the integral on the left half-circle. The poles lie at $s = -p$ and $s = -n-p-\frac 12$ for $n = 0,1,2,\ldots$, and yield 
\begin{equation}\begin{split}
I_p(x) &\sim \frac 1{2p} -\frac1\pi x^{-1} -\sum_{n=1}^\infty \frac{2(-1)^n\Gamma(n+\frac12)\Gamma(n+p+\frac12)\zeta(2n+1)}{\pi^{\frac32+2n}x^{2n+1}\Gamma(n)\Gamma(p-n+\frac12)}\\
&= \frac 1{2p} -\frac1\pi x^{-1} + \frac{4p^2-1}{4\pi^3}\zeta(3)x^{-3} -\frac{3(9-40p^2+16p^4)}{32\pi^5}\zeta(5)x^{-5}+ \cdots \,.
\end{split}\end{equation}

\bibliographystyle{ssg}
\bibliography{SYM}

\end{document}